\documentclass[journal]{IEEEtran}
\def\baselinestretch{1.2}

\usepackage{tikz}
\graphicspath{{images/}}
\usepackage{amsmath}
\usepackage{latexsym, amssymb}
\usepackage{graphicx}
\usepackage{amsmath, amsbsy}
\usepackage{amsopn, amstext}
\usepackage{ifpdf,hyperref}
\usepackage{cancel, color}
\usepackage{epstopdf}

\def\lvtimes{\vec{\ltimes}}
\def\rvtimes{\vec{\rtimes}}

\def\J{{\bf 1}}

\DeclareMathOperator{\Span}{Span}
\DeclareMathOperator{\Col}{Col}
\DeclareMathOperator{\Row}{Row}

\DeclareMathOperator{\lcm}{lcm}

\def\cal{\mathcal}

\def\diag{diag}
\def\ra{\rightarrow}

\def\lra{\leftrightarrow}

\def\a{\alpha}
\def\b{\beta}
\def\d{\delta}
\def\e{\epsilon}
\def\D{\Delta}

\def\0{{\bf 0}}

\def\argmin{argmin}

\def\A{\langle A \rangle}
\def\B{\langle B \rangle}
\def\spark{spark}
\newcommand{\R}{{\mathbb R}}
\newcommand{\Q}{{\mathbb Q}}

\newcommand{\Z}{{\mathbb Z}}

\def\dsum{\mathop{\sum}\limits}

\newtheorem{thm}{Theorem}[section]
\newtheorem{dfn}[thm]{Definition}
\newtheorem{prp}[thm]{Proposition}
\newtheorem{exa}[thm]{Example}
\newtheorem{lem}[thm]{Lemma}
\newtheorem{cor}[thm]{Corollary}
\newtheorem{rem}[thm]{Remark}
\newtheorem{alg}[thm]{Algorithm}

\begin{document}

\title{ From Signal Space To  STP-CS}
\author{Daizhan Cheng
	\thanks{This work is supported partly by the National Natural Science Foundation of China (NSFC) under Grants 62073315.}
}

\maketitle

\begin{abstract}
Under the assumption that a finite signal with different sampling lengths or different sampling frequencies is considered as equivalent signals, the signal space is considered as the quotient space of $\R^{\infty}$ over equivalence. The topological structure and the properties of signal space are investigated. Using them some characteristics of semi-tensor product based compressed sensing (STP-CS) are revealed. Finally, a systematic analysis of the construction of sensing matrix based on balanced incomplete block design (BIBD) is presented.
\end{abstract}


\IEEEpeerreviewmaketitle

\section{Introduction}

The compressed sensing (CS), as a new technique for signal processing, was initially proposed by Donoho, Candes, Tao et al. in 2006 \cite{don06,can06}.  The basic idea of CS is to compress a signal $x$ of dimension $n$ through a matrix $A\in {\cal M}_{m\times n}$, $m<<n$, called the sensing matrix, to get a sampled data $y\in R^m$. That is,
\begin{align}\label{1.1}
y=Ax.
\end{align}

The sensing matrix is obtained as follows:
Assume the original signal is $\theta\in \R^n$, and the sampled data is $y=\Phi \theta\in \R^m$. Then CS tries to design $\Psi$ such that $\theta=\Psi x$. where $x$ is with limited number nonzero entries. Then the sensing matrix is obtained as $A=\Phi \Psi$.
Since $x$ is constructed in such a way, it is possible to recover $x$ from sampled date $y$ \cite{can06b,can07}.

 A fundamental characteristic is the spark of $A$, which is the smallest number of the dependent columns of $A$. Then the follows is a criterion to justify if the original signal can be  recovered from sampled data.

Denote by $w(x)$ the number of nonzero entries in vector $x$,  and $\Sigma^n_k$ the set of $x\in \R^n$ with $w(x)\leq k$. Then we have the following result.

\begin{prp}\label{p1.1} \cite{don03} Consider equation (\ref{1.1}). Assume
 $\spark(A)>2k$, then the equation has at most one solution $x\in \Sigma^n_k$.
 \end{prp}
 Hence the CS broken the
Nyquist sampling theorem, which says that to avoid aliasing the sample frequency should be greater than twice of signal highest frequency \cite{ham77}.

Construct the sensing matrix is one of the fundamental issues in CS. It has been investigated by many authors. For instance, \cite{do12,cle14} proposed a technique to construct a structurally random matrix for the sensing matrix. Many deterministic sensing matrices have been constructed \cite{ami11,xu14,yua15}. To reduce the storage space of the sensing matrices, many methods have been developed, including orthogonal bases \cite{dua12}, low-rank matrices \cite{lee13,cai14}, etc.

Semi-tensor product (STP) of matrices/vectors is a generalization of conventional matrix/vector product, which removes the dimension restriction on factor matrices/vectors of the conventional one\cite{che12}. The STP has firstly been used to CS by the pioneer work of \cite{xie16}, where the technique is named as STP-CS. The basic idea for STP-CS is to replace fundamental CS-model (\ref{1.1}) by
\begin{align}\label{1.2}
y=A\ltimes x (=(A_0\otimes I_s)x),
\end{align}
where $A_0\in {\cal M}_{m_0\times n_0}$,  $m=m_0s$ and $n=n_0s$.

There are two obvious advantages of STP-CS: (i) It can reduce the storage of sensing matrix; (ii) it can be used for all $n$ dimensional signals as long as $m|n$.

There are some fundamental characteristics which are used to evaluate the quality of sensing matrices.

\begin{dfn}\label{d1.101} \cite{dua11}  (RIP(restricted isometry property))
Matrix
 $A$ is said to satisfy the $(k,\d)$-RIP, if there exists  $\d=\d_k^A\in (0,1)$, such that
\begin{align}\label{1.3}
(1-\d)\|x\|_2^2\leq \|Ax\|_2^2\leq (1+\d)\|x\|_2^2,\quad \forall x\in \Sigma_k.
\end{align}
\end{dfn}

When $k<m< n$,  RIP can ensure a lossless recovering of the signal.

Unfortunately, in general, verifying RIP is difficult.  An alternative criterion is the coherence.

\begin{dfn}\label{d1.2} \cite{dua11}

Assume $A\in {\cal M}_{m\times n}$, $m<n$. the coherence of $A$, denoted by  $\mu(A)$, is defined as follows.
\begin{align}\label{1.4}
\mu(A)=\max_{1\leq i\neq j\leq n}\frac{<a_i,a_j>}{\|a_i\|\|a_j\|},
\end{align}
where, $a_i=\Col_i(A)$.
\end{dfn}

\begin{rem}\label{r1.3}
\begin{itemize}
\item[(i)] \cite{xie16}
\begin{align}\label{1.5}
\mu(A)\in \left[ \sqrt{\frac{n-m}{m(n-1)}},1\right].
\end{align}
\item[(ii)] A sufficient condition for a lossless recovering the signal is \cite{dua11}
\begin{align}\label{1.6}
k<\frac{1}{2}\left(1+\frac{1}{\mu(A)}\right).
\end{align}
Hence the smaller the $\mu(A)$ the larger the $k$ is allowed, which means more signals can be treated.
\end{itemize}
\end{rem}

It was proved in \cite{xie16} that  the sensing matrix $A\otimes I_s$ has exactly the same CS-related properties as $A$.

\begin{prp}\label{p1.4}\cite{xie16} Consider $A$ and $A\otimes I_s$, where $A\in {\cal M}_{m\times n}$, $m<n$, and $s\in \Z^+$.
\begin{itemize}
\item[(i)]
\begin{align}\label{1.7}
\spark(A\otimes I_s)=\spark(A).
\end{align}
\item[(ii)] $A$ satisfies $(k,\d)$-RIP, if and only, $A\otimes I_s$ satisfies  $(k,\d)$-RIP with the same $k$ and $\d$. (The ``only if" part is claimed and proved in \cite{xie16}, and the ``if" part is straightforward verifiable.)
\item[(iii)]
\begin{align}\label{1.8}
\mu(A\otimes I_s)=\mu(A).
\end{align}
\end{itemize}
\end{prp}

Proposition \ref{p1.4} provides a theoretical foundation for STP-CS.

Since then, there have been considerable researches for STP-CS.  For instance, a further development of STP-CS called the $PTP-CS$ is presented by \cite{pen19}, \cite{lia22} considered how to construct a random sensing matrix for STP-CS.    Some applications shown the efficiency of  STP-CS. For instance,   the design of one-bit compressed sensing by STP-CS was proposed by \cite{hou23},
\cite{cha23} uses STP-CS for secure medical image transmission;  the problem of reducing storage space of sensing matrix using STP-CS was investigated by \cite{wan17}; a secure image encryption scheme by using STP-CS is proposed by  \cite{wen20}; \cite{jos19} considered the storage constrained smart meter sensing using STP-CS; \cite{ye23} proposed a reversible image hiding algorithm using STP-CS, etc.

The purpose of this paper is threefold: (1) Propose the structure of signal space. under the assumption that a finite signal with different sampling lengths or different sampling frequencies is considered as equivalent signals, the signal space is considered as the quotient space of $\R^{\infty}$ over equivalence. The structure of signal space provides a solid framework for design of STP-CS. (2) Reveal some fundamental properties of STP-CS. (3) Propose a new technique for constructing sensing matrix based on BIBD.

The rest of this paper is organized as follows.
Section 2 is a systematic review on both left and right STPs, their quotient spaces, and their vector space structure. Section 3 investigates the topological structure of the signal space, which is the quotient space of cross-dimensional Euclidian space, and its topology is deduced from distance. In Section 4 the signal space is embedded into various functional spaces, including Fr\.{e}chet space, Balach space, and Hilbert space. These functional spaces provide different structures for signal space, which reveals related properties of signal space. Section   5 provides a generalized dimension free STP-CS, which contains the STP-CS in current literature as its spacial case. An precise lossless recovering condition for STP-CS is obtained. Section 6 proposes a new constructing technique for balanced incomplete block design (BIBD) based design of sensing matrix. Finally, Section 7 is a brief conclusion. 

Before ending this section we give a list of notations.

\begin{itemize}

\item $\R$: The set of real numbers.

\item $\Q$ ($Q^+$) : The set of (positive) rational numbers.

\item $\Z$ ($\Z^+$): Set of (positive) integers.

\item $t|r$ ($t \nmid r$): $t,r\in \Z^+$ and $r/t\in \Z^+$ ($ r/t\not\in \Z^+$).

\item $\R^{\infty}$: $\R^{\infty}=\bigcup_{n=1}^{\infty}\R^n$.

\item ${\cal M}_{m\times n}$: set of $m\times n$ real matrices.

\item ${\cal M}_{\mu}:=\left\{A_{m\times n}\;|\; m/n=\mu\right\} $ where $\mu\in \Q^+$.

\item ${\cal M}:=\bigcup_{m=1}^{\infty} \bigcup_{n=1}^{\infty} {\cal M}_{m\times n}$

\item $[a,b]$: the set of integers $\{a,a+1,\cdots,b\}$, where $a\leq b$.

\item $\lceil x\rceil$: the smallest integer upper bound of $x$, i.e., the smallest integer $s\geq x$.

\item $\lcm(n,p)$  : the least common multiple of $n$ and $p$.

\item$\gcd(n,p)$ : the greatest common divisor of $n$ and $p$.

\item $\d_n^i$: the $i$-th column of the identity matrix $I_n$.

\item $\D_n:=\left\{\d_n^i\vert i=1,\cdots,n\right\}$.

\item $\J_k:=(\underbrace{1,\cdots,1}_k)^T$.

\item $\Span(\cdot)$: subspace or dual subspace generated by $\cdot$.

\item $\bar{+}$: semi-tensor addition.

\item $\ltimes$: (left) matrix-matrix semi-tensor product, which is the default one.

\item $\rtimes$: right semi-tensor product of matrices.

\item $\lvtimes$ ($\rvtimes$): left (right) matrix-vector semi-tensor product.

\item $\Sigma^n_k$: set of $n$ dimensional vector with less than or equal to $k$ non-zero entries.

\item $\Sigma^n_{k/s}$: set of $n$ dimensional vector with less than or equal to $k$ non-zero entries per $s$ length.

\item $w(x)=\ell_0(x)$: nonzero entries of vector $x$(called the degree of $x$).

\item $\spark(A)$: the smallest number of dependent columns of $A$.

\item $\mu(A)$: coherence of $A$.
\end{itemize}

\section{Left and Right STP - Preliminaries}

\begin{dfn}\label{d2.1}\cite{che12} Assume $A\in {\cal M}_{m\times n}$, $B\in {\cal M}_{p\times q}$, $x\in \R^r$, $t=\lcm(n,p)$, and $s=\lcm(n,r)$.
\begin{itemize}
\item[(i)] The left matrix-matrix (MM-) STP is defined by
\begin{align}\label{2.1}
A\ltimes B:=\left(A\otimes I_{t/n}\right)\left(B\otimes I_{t/p}\right)\in {\cal M}_{mt/n\times qt/p}.
\end{align}
\item[(ii)] The right MM-STP is defined by
\begin{align}\label{2.2}
A\rtimes B:=\left( I_{t/n}\otimes A\right)\left(I_{t/p}\otimes B\right)\in {\cal M}_{mt/n\times qt/p}.
\end{align}
\item[(iii)] The left matrix-vector (MV-) STP is defined by
\begin{align}\label{2.3}
A\lvtimes x:=\left(A\otimes I_{s/n}\right)\left(x\otimes \J_{s/r}\right)\in \R^s.
\end{align}
\item[(iv)] The right MV-STP is defined by
\begin{align}\label{2.4}
A\rvtimes x:=\left( I_{s/n}\otimes A\right)\left(\J_{s/r}\otimes x\right)\in \R^s.
\end{align}
\end{itemize}
\end{dfn}

Denote the set of finite-dimensional matrices by
$$
{\cal M}=\bigcup_{m=1}^{\infty}\bigcup_{n=1}^{\infty}{\cal M}_{m\times n}.
$$
And the set of finite-dimensional vectors by
$$
\R^{\infty}=\bigcup_{n=1}^{\infty}\R^n.
$$

The basic properties of STPs are listed as follows:

\begin{prp}\label{p2.2}\cite{che12} Let $A,B,C\in {\cal M}$, $x,y\in \R^{\infty}$.
\begin{enumerate}
\item[(1)] (Consistency)
\begin{itemize}
\item[(i)] When $n=p$, the MM-STPs are degenerated to conventional MM-product. That is,
\begin{align}\label{2.5}
A\ltimes B=A\rtimes B=AB.
\end{align}
\item[(ii)] When $n=r$, the MV-STPs are degenerated to conventional MV-product. That is,
\begin{align}\label{2.6}
A\lvtimes x=A\rvtimes x=Ax.
\end{align}
\end{itemize}
\item[(2)] (Associativity)
\begin{itemize}
\item[(i)]
\begin{align}\label{2.7}
(A\ltimes B)\ltimes C=A\ltimes (B\ltimes C).
\end{align}
\item[(ii)]
\begin{align}\label{2.8}
(A\rtimes B)\rtimes C=A\rtimes (B\rtimes C).
\end{align}
\item[(iii)]
\begin{align}\label{2.9}
(A\ltimes B)\lvtimes x:=A\lvtimes (B\lvtimes x).
\end{align}
\item[(iv)]
\begin{align}\label{2.10}
(A\rtimes B)\rvtimes x:=A\rvtimes (B\rvtimes x).
\end{align}
\end{itemize}
\item[(3)] (Distributivity)
\begin{itemize}
\item[(i)]
\begin{align}\label{2.11}
\begin{array}{l}
(A + B)\ltimes C=A\ltimes C+B\ltimes C,\\
C\ltimes (A + B)=C\ltimes A+C\ltimes B.
\end{array}
\end{align}
\item[(ii)]
\begin{align}\label{2.12}
\begin{array}{l}
(A + B)\rtimes C=A\rtimes C+B\rtimes C,\\
C\rtimes (A + B)=C\rtimes A+C\rtimes B.
\end{array}
\end{align}
\item[(iii)]
\begin{align}\label{2.13}
\begin{array}{l}
(A+B)\lvtimes x=A\lvtimes x+B\lvtimes x,\\
A\lvtimes (x+y)=A\lvtimes x+A\lvtimes y.\\
\end{array}
\end{align}
\item[(iv)]
\begin{align}\label{2.14}
\begin{array}{l}
(A+B)\rvtimes x=A\rvtimes x+B\rvtimes x,\\
A\rvtimes (x+y)=A\rvtimes x+A\rvtimes y.\\
\end{array}
\end{align}
\end{itemize}
\end{enumerate}
\end{prp}

Note that when the addition of two matrices (vectors) appears to the equalities in Proposition \ref{p2.2}, the dimensions of two adding members must be the same.

\begin{dfn}\label{d2.3}
\begin{itemize}
\item[(i)] Two matrices $A$ and $B$ are said to be left equivalent, denoted by $A\sim_{\ell} B$, if there exist two identities $I_{\a}$ and $I_{\b}$ such that
\begin{align}\label{2.15}
A\otimes I_{\a}=B\otimes I_{\b}.
\end{align}
The left equivalence class of $A$ is denoted by
$$
\langle A\rangle_{\ell}:=\{B\;|\; B\sim_{\ell} A\}.
$$
\item[(ii)] Two matrices $A$ and $B$ are said to be right equivalent, denoted by $A\sim_r B$, it there exists two identities $I_{\a}$ and $I_{\b}$ such that
\begin{align}\label{2.16}
I_{\a}\otimes A=I_{\b}\otimes B.
\end{align}
The right equivalence class of $A$ is denoted by
$$
\langle A\rangle_r:=\{B\;|\; B\sim_r A\}.
$$
\item[(iii)] Two vectors $x$ and $y$ are said to be left equivalent, denoted by $x\lra_{\ell} y$, if there exist two 1-vectors  $\J_{\a}$ and $\J_{\b}$ such that
\begin{align}\label{2.17}
x\otimes \J_{\a}=y\otimes \J_{\b}.
\end{align}
The left equivalence class of $x$ is denoted by
$$
\bar{x}_{\ell}:=\{y\;|\; y\lra_{\ell} x\}.
$$
\item[(iv)] Two vectors $x$ and $y$ are said to be right equivalent, denoted by $x\lra_r y$, if there exist two 1-vectors  $\J_{\a}$ and $\J_{\b}$ such that
\begin{align}\label{2.18}
\J_{\a}\otimes x=\J_{\b}\otimes y.
\end{align}
The right equivalence class of $x$ is denoted by
$$
\bar{x}_r:=\{y\;|\; y\lra_r x\}.
$$
\end{itemize}
\end{dfn}

It is easy to verify that the equivalences defined in Definition \ref{d2.3} are equivalence relations.

\begin{prp}\label{p2.4}  \cite{che16}
\begin{itemize}
\item[(i)] If $A\sim_{\ell} B$, then there exists a $C\in {\cal M}$ such that
\begin{align}\label{2.19}
A=C\otimes I_{b},\quad B=C\otimes I_{a}.
\end{align}
w.l.g. (without lose of generality) we can assume in (\ref{2.15}) the $\gcd(\a,\b)=1$, then in (\ref{2.19})
$$
b=\b, ~\mbox{and}~a=\a.
$$
\item[(ii)] If $A\sim_r B$, then there exists a $C\in {\cal M}$ such that
\begin{align}\label{2.20}
A= I_{b}\otimes C,\quad B= I_{a}\otimes C.
\end{align}
Assume in (\ref{2.16}) the $\gcd(\a,\b)=1$, then in (\ref{2.20})
$$
b=\b, ~\mbox{and}~a=\a.
$$
\item[(iii)] If $x\lra_{\ell} y$, then there exists a $z\in \R^{\infty}$ such that
\begin{align}\label{2.21}
x=z\otimes \J_{b},\quad y=z\otimes \J_{a}.
\end{align}
Assume in (\ref{2.17}) the $\gcd(\a,\b)=1$, then in (\ref{2.21})
$$
b=\b, ~\mbox{and}~a=\a.
$$
\item[(iv)] If $x\lra_r y$, then there exists a $z\in \R^{\infty}$ such that
\begin{align}\label{2.22}
x=\J_{b}\otimes z,\quad  y=\J_{a}\otimes z.
\end{align}
Assume in (\ref{2.18}) the $\gcd(\a,\b)=1$, then in (\ref{2.22})
$$
b=\b, ~\mbox{and}~a=\a.
$$
\end{itemize}
\end{prp}

\begin{dfn}\label{d2.5}
\begin{itemize}
\item[(i)] $A\in {\cal M}$ is said to be left (right) reducible, if there exist an identity matrix $I_s$, $s>1$ and an $A_0\in {\cal M}$, such that
\begin{align}\label{2.23}
A=A_0\otimes I_s, (\mbox{correspondingly,}~A=I_s\otimes A_0).
\end{align}
Otherwise, it is called left (right) irreducible.

\item[(ii)] $A\in \R^{\infty}$ is said to be left (right) reducible, if there exist an 1-vector  $\J_s$, $s>1$ and an $x_0\in \R^{\infty}$, such that
\begin{align}\label{2.24}
x=x_0\otimes \J_s, (\mbox{correspondingly,}~x=\J_s\otimes x_0).
\end{align}
Otherwise, it is called left (right) irreducible.
\end{itemize}
\end{dfn}

\begin{prp}\label{p2.6}\cite{che16}
\begin{itemize}
\item[(i)] Consider $\langle A\rangle_{\ell}$  ($\langle A\rangle_r$).  There exists a unique left irreducible $A_0^{\ell}$ (correspondingly, right irreducible $A_0^r$), such that
\begin{align}\label{2.25}
\begin{array}{l}
\langle A\rangle_{\ell}=\{A_n=A^{\ell}_0\otimes I_n\;|\; n\in \Z^+\}.\\
(\mbox{correspondingly,}~\langle A\rangle_r=\{A_n=I_n\otimes A^r_0\;|\; n\in \Z^+\}.)\\
\end{array}
\end{align}
\item[(ii)] Consider $\bar{x}_{\ell}$ ($\bar{x}_r$). There exists a unique left irreducible $x_0^{\ell}$ (correspondingly, right irreducible $x_0^r$), such that
\begin{align}\label{2.26}
\begin{array}{l}
\bar{x}_{\ell}=\{x_n=x^{\ell}_0\otimes \J_n\;|\; n\in \Z^+\}.\\
(\mbox{correspondingly,}~\bar{x}_r=\{x_n=\J_n\otimes x^r_0\;|\; n\in \Z^+\}.)\\
\end{array}
\end{align}
\end{itemize}
\end{prp}

\begin{dfn}\label{d2.7}
\begin{itemize}
\item[(i)] Consider ${\cal M}$. The left equivalent classes form the left quotient space as
\begin{align}\label{2.27}
\Sigma_{\ell}:={\cal M}/\sim_{\ell}=\{\A_{\ell}\;|\; A\in {\cal M}\}.
\end{align}
\item[(ii)] Consider ${\cal M}$. The right equivalent classes form the right quotient space as
\begin{align}\label{2.28}
\Sigma_r:={\cal M}/\sim_r=\{\A_r\;|\; A\in {\cal M}\}.
\end{align}
\item[(iii)] Consider $\R^{\infty}$. The left equivalent classes form the left quotient space as
\begin{align}\label{2.29}
\Omega_{\ell}:=\R^{\infty}/\lra_{\ell}=\{\bar{x}_{\ell}\;|\; x\in \R^{\infty}\}.
\end{align}
\item[(iv)] Consider $\R^{\infty}$. The right equivalent classes form the right quotient space as
\begin{align}\label{2.30}
\Omega_r:=\R^{\infty}/\lra_r=\{\bar{x}_r\;|\; x\in \R^{\infty}\}.
\end{align}
\end{itemize}
\end{dfn}

\begin{rem}\label{r2.7} In this section we carefully reviewed the left(right) STP, left(right) equivalence, and left(right) quotient space for both matrices and vectors. It is easy to see that the left (STP) system, including the  left STP, left equivalence, and left quotient space, is ``mirror symmetric" to the right (STP) system, including  the  right STP, right equivalence, and right quotient space. Up to this paper, all the researches are concentrated mainly on the left system. In the rest of this paper, we are only interested on left system too. We claim that all the results obtained for the left system are also available for the right system. Unless elsewhere stated, the corresponding verifications are straightforward.
\end{rem}

Hereafter we assume the default system is the left one. That is,
\begin{align}\label{2.31}
\begin{cases}
\sim:=\sim_{\ell},\\
\lra:=\lra_{\ell},\\
\A:=\A_{\ell},\\
\bar{x}:=\bar{x}_{\ell},\\
\Omega:=\Omega_{\ell},\\
\Sigma:=\Sigma_{\ell}.\\
\end{cases}
\end{align}
Unless elsewhere stated.

\section{Signal Space}

A signal  length $n$ can be considered as a vector in Euclidian space $\R^n$ \cite{dua11}.

\begin{dfn}\label{d3.0.1}
Two signals $x\in \R^m$ and $y\in \R^n$ are said to be equivalent if $x\lra y$.
\end{dfn}

\begin{rem}\label{r3.0.2}

\begin{itemize}
\item[(i)]
According to Proposition \ref{p2.4}, $x\lra_r y$ means there exists a $z$ such that $x=\J_{\a}\otimes z$ and $y=\J_{\b}\otimes z$. Hence we have
$$
\begin{array}{l}
x=\underbrace{(z^T,z^T,\cdots,z^T)^T}_{\a},\\
y=\underbrace{(z^T,z^T,\cdots,z^T)^T}_{\b}.\\
\end{array}
$$
That is, $x$ and $y$ are obtained from same signal $z$ with different sampling time lengths. This fact makes the physical meaning of equivalence clear.

If we consider $x$ and $y$ from frequency domain, they are exactly the same.

\item[(ii)] Similarly, $x\lra_{\ell} y$ means there exists a $z$ such that $x=z\otimes \J_{\a}$ and $y=z\otimes \J_{\b}$. Assume $z=(z_1,z_2,\cdots,z_k)^T$, then  we have
$$
\begin{array}{l}
x=(\underbrace{z_1,z_1,\cdots,z_1}_{\a},  \underbrace{z_2,z_2,\cdots,z_2}_{\a},\cdots,\underbrace{z_k,z_k,\cdots,z_k}_{\a})^T,\\
y=(\underbrace{z_1,z_1,\cdots,z_1}_{\b},  \underbrace{z_2,z_2,\cdots,z_2}_{\b},\cdots,\underbrace{z_k,z_k,\cdots,z_k}_{\b})^T .\\
\end{array}
$$
It can  be considered as both $x$ and $y$ are obtained from same signal $z$ with different sampling frequencies (with certain approximation).
\end{itemize}
Hence, from finite signal point of view, the right (or left) equivalence is physically meaningful.
\end{rem}

It is clear that under the left equivalence the signal space becomes
$$
\Omega=\Omega_{\ell}.
$$
And a signal $x$ becomes $\bar{x}$. Let $x_0\in \bar{x}$ be irreducible, then $x_0$ is called the atom signal, and we define
\begin{align}\label{3.0.1}
\dim(\bar{x}):=\dim(x_0).
\end{align}

\subsection{Vector Space Structure on Signal Space}

\begin{dfn}\label{d3.1.1}\cite{che16} Let $x,y\in \R^{\infty}$ with $x\in \R^m$ and $y\in \R^n$, $t=\lcm(m, n)$. Then the left (right) semi-tensor addition (STA) of $x$ and $y$ is defied by
\begin{align}\label{3.1.1}
\begin{array}{l}
x\vec{\pm}_{\ell}y:=\left(x\otimes \J_{t/m}\right)\pm\left(y\otimes \J_{t/n}\right)\in \R^t;\\
x\vec{\pm}_{r}y:=\left(\J_{t/m}\otimes x\right)\pm\left(\J_{t/n}\otimes y\right)\in \R^t.
\end{array}
\end{align}
\end{dfn}

The physical meaning of these additions are very clear: Assume $x$ is a signal with period $m$ and $y$ is a signal with period $n$, then the addition of these two signals, $x\vec{+}_r y$,  is a signal with period $t$. Similarly, assume $x$ is a signal with frequency $1/m$ and $y$ is a signal with period $1/n$, then the addition of these two signals, $x\vec{+}_{\ell} y$, is a signal with frequency $1/t$.

\begin{figure}
\centering
\setlength{\unitlength}{5mm}
\begin{picture}(16,9)(-1,-1)\thicklines
\put(-1,0){\vector(1,0){14}}
\put(0.1,-1){\vector(0,1){8}}
\put(0.2,-0.8){0}
\put(11.8,-0.8){12}
\put(5.8,-0.8){6}
\put(2,-0.8){2}
\put(3,-0.8){3}
\put(13.2,3){$x$}
\put(13.2,1){$y$}
\put(3.1,7){\vector(-1,-1){1}}
\put(2.5,7.2){$x\vec{+}_r y$}
\put(6,7){\vector(-1,-1){1}}
\put(5.5,7.2){$x\vec{+}_{\ell} y$}
\put(0,2){\line(1,-1){1}}
\put(1,1){\line(1,1){1}}
\put(2,2){\line(1,-1){1}}
\put(3,1){\line(1,1){1}}
\put(4,2){\line(1,-1){1}}
\put(5,1){\line(1,1){1}}
\put(6,2){\line(1,-1){1}}
\put(7,1){\line(1,1){1}}
\put(8,2){\line(1,-1){1}}
\put(9,1){\line(1,1){1}}
\put(10,2){\line(1,-1){1}}
\put(11,1){\line(1,1){1}}
\put(12,2){\line(1,-1){1}}
{\color{red}
\put(0,2){\line(1,1){2}}
\put(2,4){\line(1,-2){1}}
\put(3,2){\line(1,1){2}}
\put(5,4){\line(1,-2){1}}
\put(6,2){\line(1,1){2}}
\put(8,4){\line(1,-2){1}}
\put(9,2){\line(1,1){2}}
\put(11,4){\line(1,-2){1}}
\put(12,2){\line(1,1){1}}
}
{\color{blue}
\put(0,4){\line(1,0){1}}
\put(1,4){\line(1,2){1}}
\put(2,6){\line(1,-3){1}}
\put(3,3){\line(1,2){1}}
\put(4,5){\line(1,0){1}}
\put(5,5){\line(1,-1){1}}
\put(6,4){\line(1,0){1}}
\put(7,4){\line(1,2){1}}
\put(8,6){\line(1,-3){1}}
\put(9,3){\line(1,2){1}}
\put(10,5){\line(1,0){1}}
\put(11,5){\line(1,-1){1}}
\put(12,4){\line(1,0){1}}
}
{\color{green}
\put(0,4){\line(1,0){3}}
\put(3,4){\line(1,2){1}}
\put(4,6){\line(1,0){1}}
\put(5,6){\line(1,-3){1}}
\put(6,3){\line(1,0){1}}
\put(7,3){\line(1,2){1}}
\put(8,5){\line(1,0){3}}
\put(11,5){\line(1,-1){1}}
\put(12,4){\line(1,0){1}}
}
\thinlines
\put(11.8,0){\line(0,1){4}}
\put(5.8,0){\line(0,1){4}}
\put(1.8,0){\line(0,1){2}}
\put(2.8,0){\line(0,1){2}}
\end{picture}
\caption{Signal Addition \label{Fig3.1}}
\end{figure}
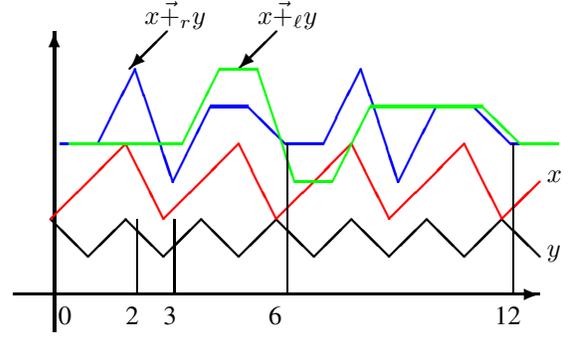

Figure \ref{Fig3.1} shows that the black signal has period $2$ and the red signal has period $3$, then their addition $x\vec{+}_r y$, (blue line) is a signal with period $\lcm(2,3)=6$; their addition $x\vec{+}_{\ell} y$, (green line) is a signal with frequency $1/6$.

\begin{prp}\label{p3.1.2}\cite{che16} The addition defined by (\ref{3.1.1}) ls consistent with the equivalence. That is, if $x_1\lra x_2$ and $y_1\lra y_2$, then
\begin{align}\label{3.1.2}
x_1\vec{\pm}y_1\lra x_2\vec{\pm} y_2.
\end{align}
\end{prp}

Using Proposition \ref{p3.1.2}, the following results are obvious.

\begin{prp}\label{p3.1.3}\cite{che16}
\begin{itemize}
\item[(i)] $\vec{+}$ is properly defined on $\Omega$ by
\begin{align}\label{3.1.3}
\bar{x}\vec{\pm}\bar{y}=\overline{x\vec{\pm} y},\quad \bar{x},\bar{y}\in \Omega.
\end{align}
\item[(ii)] Define the scalar product on $\Omega$ by
\begin{align}\label{3.1.4}
a\cdot \bar{x}:=\overline{ax},\quad a\in \R, \bar{x}\in \Omega.
\end{align}
With $\vec{+}$ and scalar product $\cdot$, $\Omega$ is a vector space.
\end{itemize}
\end{prp}

\subsection{Topological Structure on $\Omega$}

\begin{dfn}\label{d3.2.1}  Let $x,y\in \R^{\infty}$ with $x\in \R^m$ and $y\in \R^n$, $t=\lcm(m, n)$.
\begin{itemize}
\item[(i)]
The inner product of $x$ and $y$ is defined by
\begin{align}\label{3.2.1}
\langle x,y\rangle_{{\cal V}}:=\frac{1}{t} \langle (x\otimes \J_{t/m},y\otimes \J_{t/n} \rangle,
\end{align}
where
$$
\langle (x\otimes \J_{t/m},y\otimes \J_{t/n}\rangle=(x\otimes \J_{t/m})^T(y\otimes \J_{t/n})
$$
is the conventional inner product on $\R^t$.

\item[(ii)] The norm of $x$ is defined by
\begin{align}\label{3.2.2}
\|x\|_{{\cal V}}:=\sqrt{\langle x,x\rangle_{{\cal V}}}=\frac{1}{\sqrt{n}}\sqrt{x^Tx}.
\end{align}
\item[(iii)] The distance of $x$ and $y$ is defined by
\begin{align}\label{3.2.3}
d_{{\cal V}}(x,y):= \|x\vec{-}y\|_{{\cal V}}.
\end{align}
\end{itemize}
\end{dfn}

Using this inner product, the angle between two elements  $x,y\in \R^{\infty}$, denoted by $\theta=\theta_{x,y}$, is defined by
\begin{align}\label{3.2.4}
\cos(\theta)=\frac{\langle x,y\rangle_{{\cal V}}}{\|x\|_{{\cal V}}\|y\|_{{\cal V}}}.
\end{align}

Using this distance, the distance deduced topology can be obtained, which is denoted by ${\cal T}_d$. Apart from this ${\cal T}_d$, there is another topology on $\R^{\infty}$. Naturally, each $\R^n$ can be considered as a component (i.e., a clopen set) of $\R^{\infty}$. And within each $\R^n$ the conventional Euclidian space topology is used. Then overall, such a topology is called the natural topology, denoted by ${\cal T}_n$. $(\R^{\infty},{\cal T}_n)$ is disconnected. Precisely speaking, its fundamental group is $\Z^+$.

It is easy to verify the  following result.

\begin{prp}\label{p3.2.2}\cite{che16} Consider $\R^{\infty}$. Let $x,y\in \R^{\infty}$. Then
$d(x,y)=0$, if and only if, $x\lra y$.
\end{prp}


\begin{prp}\label{p3.2.3} \cite{che16} The inner product defined by (\ref{3.2.1}) is consistent with the equivalence. That is,
if $x_1\lra x_2$ and $y_1\lra y_2$, then
\begin{align}\label{3.2.4}
\langle x_1,y_1\rangle_{{\cal V}}=\langle x_2,y_2\rangle_{{\cal V}}.
\end{align}
\end{prp}

According to Proposition \ref{p3.2.3}, the inner product, norm, distance defined by Definition \ref{d3.2.1} for $\R^{\infty}$ can be extended to $\Omega$.
Hence the following definitions for $\Omega$ are all properly defined.

\begin{dfn}\label{d3.2.4}
\begin{itemize}
\item[(i)]
The inner product of $\bar{x}$ and $\bar{y}$ is defined by
\begin{align}\label{3.2.5}
\langle \bar{x},\bar{y}\rangle_{{\cal V}}:=\langle x, y\rangle_{{\cal V}}, \quad \bar{x},\bar{y}\in \Omega.
\end{align}

\item[(ii)] The norm of $\bar{x}$ is defined by
\begin{align}\label{3.2.6}
\|\bar{x}\|_{{\cal V}}:=\|x\|_{{\cal V}}, \bar{x}\in \Omega.
\end{align}
\item[(iii)] The distance of $\bar{x}$ and $\bar{y}$ is defined by
\begin{align}\label{3.2.7}
d_{{\cal V}}(\bar{x},\bar{y}):= d_{{\cal V}}(x,y),  \bar{x},\bar{y}\in \Omega.
\end{align}
\end{itemize}
\end{dfn}

With the topology deduced by (\ref{3.2.7}) $\Omega$ is a topological space. Because of Proposition (\ref{3.2.3}), this distance deduced topology is homeomorphic to quotient topology. That is,
$$
\Omega=\left[\left(\R^{\infty},{\cal T}_n\right)/\lra\right]\cong \left(\R^{\infty},{\cal T}_d\right).
$$

In the following proposition we summarize some known basic properties of $\Omega$.

\begin{prp}\label{p3.2.5}\cite{che16,che19} Consider the signal space $\Omega$.
\begin{itemize}
\item(i)] $\Omega$ is an infinite dimensional vector space, while each element has finite dimension. That is,
$$
\dim(\bar{x})<\infty,\quad \bar{x}\in \Omega.
$$
\item[(ii)] As a topological space, $\Omega$ is second countable, separable, Hausdorff.
\item[(iii)] $\Omega$ is an inner product space, but not Hilbert.
\end{itemize}
\end{prp}

Since $\Omega$ is an inner product space, then the following geometric structure is obvious.

\begin{cor}\label{c3.2.6}
\begin{itemize}
\item[(i)] For any two points $\bar{x},\bar{y}\in \Omega$, the angle between them (the same as the angle between $x$ and $y$) ,   denoted by  $\theta_{\bar{x},\bar{y}}$,  is determined by
\begin{align}\label{3.2.8}
\cos(\theta_{\bar{x},\bar{y}})= \cos(\theta_{x,y})=\frac{\langle x,y\rangle_{{\cal V}}}{\|x\|_{{\cal V}}\|y\|_{{\cal V}}}.
\end{align}
\item[(ii)] $\bar{x}$ and $\bar{y}$ (as well as $x$ and $y$) are said to be parallel (orthogonal), if $\cos(\theta_{x,y})=1$ (correspondingly, $\cos(\theta_{x,y})=0$ ).
\end{itemize}
\end{cor}

Let $x\in \R^m$. The projection of $x$ onto $\R^n$, denoted by $\pi^m_n(x)$, is defined by
\begin{align}\label{3.2.9}
\pi^m_n(x)=\argmin_{y}(d_{{\cal V}}(y,x)).
\end{align}

\begin{prp}\label{p3.2.7} \cite{che19}
\begin{itemize}
\item[(i)] The projection of $x\in \R^m$ onto $\R^n$ is
\begin{align}\label{3.2.10}
y_0:=\pi^m_n(x)=\Pi^m_n\ltimes x,
\end{align}
where ($t=\lcm(m,n)$)
$$
\Pi^m_n=\frac{n}{t}\left(I_n\otimes \J^T_{t/n}\right)\left(I_m\otimes \J_{t/m}\right).
$$
\item[(ii)] $x\vec{-} y_0$ is perpendicular to $y_0$, ($(x\vec{-} y_0)\perp y_0$).  That is, the angle between them, denoted by $\theta$ satisfies
\begin{align}\label{3.2.11}
\cos(\theta)=0.
\end{align}
\item[(iii)] For any $y\in \R^n$, $(x\vec{-} y_0)\perp y$.
\end{itemize}
\end{prp}

(See Figure \ref{Fig3.2} for the projection.)

\begin{figure}
\centering
\setlength{\unitlength}{6mm}
\begin{picture}(7,4)\thicklines
\put(3,1){\vector(1,0){3}}
\put(3,1){\vector(1,1){3}}
\put(6,1){\vector(0,1){3}}
\put(6,1){\vector(-3,-1){2}}
\put(2,1.4){$\R^n$}
\put(4.5,3){$x$}
\put(6.2,2.5){$x\vec{-} y_0$}
\put(4.5,1.2){$y_0$}
\put(3.7,0.3){$y$}
\put(2.6,0.8){$0$}
\thinlines
\put(0,0){\line(1,0){6}}
\put(2,2){\line(1,0){6}}
\put(0,0){\line(1,1){2}}
\put(6,0){\line(1,1){2}}

\end{picture}
\caption{Projection \label{Fig3.2}}
\end{figure}
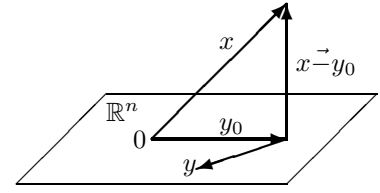
\begin{rem}\label{r3.2.8} The projection $\pi^m_n$ can be extended naturally to a projection between subspaces of $\Omega$.
\end{rem}
%

\subsection{Linear Mapping on $\Omega$}

Consider the set of matrices ${\cal M}$ and the equivalence $\sim=\sim_{\ell}$. Similarly to vectors, we define the reducibility.

\begin{dfn}\label{d3.3.1} Let $A\in {\cal M}$. $A$ is said to be (left) reducible, if there exists an identity matrix $I_s$, $s>1$ and a matrix $A_0$ such that
\begin{align}\label{3.3.1}
A= A_0 \otimes I_s.
\end{align}
Otherwise, it is irreducible.
\end{dfn}

Using Proposition \ref{p2.4}, one sees easily that in $\A$ there exists a unique irreducible $A_0$ such that
$$
\A=\{A_n=A_0\otimes I_n\;|\;n\in \Z^+\}
$$

 Next, consider the linear mapping on $\R^{\infty}$. We have the following result:

 \begin{prp}\label{p3.3.2} \cite{che16} Assume $A,B\in {\cal M}$ with $A\sim B$, and  $x,y\in \R^{\infty}$ with $x\lra y$. Then the MV-STP $\lvtimes$ is consistent with both equivalences. That is,
\begin{align}\label{3.3.2}
A\lvtimes x\lra B\lvtimes y.
\end{align}
\end{prp}

 Recall the quotient space of matrices, $\Sigma=\Sigma_{\ell}$, using Proposition \ref{p3.3.2}, the following result is obvious.

 \begin{cor}\label{c3.3.3}
\begin{itemize}
\item[(i)]  A linear mapping on image space $\Omega$ is a mapping $\pi: \Sigma\times \Omega\ra \Omega$, which can be expressed by
 \begin{align}\label{3.3.3}
\A \lvtimes \bar{x}:=\overline{A\lvtimes x}.
\end{align}
\item[(ii)] For two linear mappings $\A$ and $\B$,
 \begin{align}\label{3.3.4}
\A \lvtimes (B \lvtimes \bar{x})=(\A \ltimes \B) \lvtimes x,\quad \A,\B\in \Sigma, \bar{x}\in \Omega.
\end{align}
\end{itemize}
\end{cor}

\section{Structure of Signal Space}

\subsection{Orthonormal Basis of Signal Space}

First, we search a basis of $\Omega=\Omega_{\ell}$.

It was claimed  by \cite{zha18} that
\begin{align}\label{5.1.1}
{\cal B}_{{\cal V}}:=\left\{1,\d_i^j\;|\; j<i, \gcd(i,j)=1;\; i\in \Z^+\right\}
\end{align}
is a basis of $\Omega$. Unfortunately, ${\cal B}_{{\cal V}}$ is only a generating set of $\Omega$, which was proved by \cite{zha18}.
But there are some linearly depending terms. Say,
$$
\d_4^1\vec{+} \d_4^3\vec{-}\d_2^1=0.
$$

Searching a basis of $\Omega$ is a fundamental task for understanding and using the signal space.

Denote
$$
\begin{array}{l}
{\cal B}_{{\cal V}_n}:={\cal B}_{{\cal V}}\bigcap \D_n,\\
{\cal B}^m_{{\cal V}}:=\bigcup_{n\leq m} {\cal B}_{{\cal V}_n},\\
\Omega_1=\Span\{1\},\\
\Omega_n=\Span\{\d_n^j\;|\; j<n,\;\gcd(j,n)=1\}, \quad n\geq 2,\\
\Omega^m=\bigcup_{n\leq m} \Omega_n.\\
\end{array}
$$

$1<s\in \Z^+$ is said to be a multi-fold divisor of $n\in \Z^+$ if $s^2|n$.

Note that since $\Omega =\Span({\cal B}_{{\cal V}})$, we can gradually choose
$$
{\cal D}_{{\cal V}_n}\subset {\cal B}_{{\cal V}_n},\quad n=1,2,\cdots,
$$
and denote
$$
{\cal D}^m_{{\cal V}}:=\bigcup_{n\leq m}{\cal D}_{{\cal V}_n},
$$
such that the elements of ${\cal D}^m{{\cal V}}$ are linearly independent, and
\begin{align}\label{5.1.4}
\Omega^m=\Span({\cal D}^m{{\cal V}}).
\end{align}
That is, ${\cal D}^m_{{\cal V}}$ is a basis of $\Omega^m$.

Let $m\ra \infty$. The basis of $\Omega$ is obtained.

To describe this process clearly, the following theory shows the relationship between ${\cal D}_{{\cal V}_n}$ and ${\cal B}_{{\cal V}_n}$.

\begin{thm}\label{t5.1.1}
\begin{align}\label{5.1.405}
\Omega^{n-1}+\Span({\cal D}_{{\cal V}_n})=\Omega^n,
\end{align}
if and only if, $n$ has no multi-fold divisor.
\end{thm}

\noindent{\it Proof.}

Since
$$
\Omega^n=\Omega^{n-1}+\Span({\cal B}_{{\cal V}_n}),
$$
(\ref{5.1.405}) tells us how to choose the elements from ${\cal B}_{{\cal V}_n}$, which are linearly independent with elements in
$\Omega^{n-1}$.

(Sufficiency:)

Assume $n$ has no multi-fold divisor. We have to show that
\begin{align}\label{5.1.5}
\left\{\Omega^{n-1}, {\cal B}_{{\cal V}_n}\right\}
\end{align}
are linearly independent, which implies that ${\cal D}_{{\cal V}_n}={\cal B}_{{\cal V}_n}$.

 Let $\{\d_n^{i_1}, \cdots,\d_n^{i_s}\}$ be linearly dependent with $\Omega^{n-1}$. Then there exists a $x\in  \Omega^{n-1}$ such that it can be expressed by
\begin{align}\label{5.1.6}
\bar{x}=\overline{c_1\d_n^{i_1}\vec{+}\cdots \vec{+} c_s\d_n^{i_s}}.
\end{align}
Since
$$
c_1\d_n^{i_1}\vec{+}\cdots \vec{+} c_s\d_n^{i_s}\in \Span({\cal B}_{{\cal V}_n}),
$$ say $x_0\in \bar{x}$ is irreducible.  then $x_0\in  \Span({\cal B}_{{\cal V}_n})$. But $\dim(x_0)<n$, hence back to Euclidian space, we should have
\begin{align}\label{5.1.7}
\J_t\otimes x_0=c_1\d_n^{i_1}\vec{+}\cdots \vec{+} c_s\d_n^{i_s},
\end{align}
where $t|n$, say $n=tj$.

It is obvious that $x_0\neq 0$, because the elements in ${\cal B}_{{\cal V}_n}$ are linearly independent. Say, the $i$-th component of $x_0$ denoted by $x_0^i\neq 0$. Then
$$
\J_t\otimes x_0=\begin{bmatrix}
\d_j^i\\
\d_j^i\\
\vdots\\
\d_j^i
\end{bmatrix}
+\begin{bmatrix}
\xi_j^{-i}\\
\xi_j^{-i}i\\
\vdots\\
\xi_j^{-i}
\end{bmatrix},
$$
where $\xi_j^{-i}\in \R^j$ with $i$-th component equals $0$. To meet (\ref{5.1.7}), we need
\begin{align}\label{5.1.8}
\d_n^{i+(k-1)t},\quad \forall k\in [1,j].
\end{align}
We claim that there exists at least one $k$ such that
\begin{align}\label{5.1.9}
i+(k-1)j (\mod~ t)=0.
\end{align}

Since $n$ has no multi-fold divisor, we have
$\gcd(j,t)=1$. (Otherwise, say $\gcd(j,t)=r$, then  $n=jt$ has $r$ as its multi-fold divisor. )

Then for any $i<t$ as $k$ runs from $1$ to $t$ $i+(k-1)j (\mod~ t)$ takes all values from $[0,t-1]$. That is, there exists a $k_0\in [1,m]$ such that $i+(k_0-1)j=\mu t$. That is, $\gcd(i+(k_0-1)j,n)>0$. Hence
$$
\d_n^{i+(k_0=1)t}\not \in {\cal B}_{{\cal V}_n}.
$$
That is, (\ref{5.1.8}) does not satisfied, and hence  (\ref{5.1.7} is not true. Hance, (\ref{5.1.5}) is not a linearly independent set.

(Necessity) Assume  $n=k^2j$. where $k>1$. We have to show (\ref{5.1.5}) is a  linearly dependent set.   We can, w.l.g., assume $k$ is a prime number. Then
$$
\begin{array}{l}
\J_k\otimes \d^i_{kj}
=\d_{n^i}+\d_{n}^{i+j}+\cdots+\d_{n}^{i+(k-1)j}\\
~~~~~~~~~~~~ i\in [1,k-1].
\end{array}
$$
Note that
$$
\gcd(i+sj,n)=\gcd(i,k)=1,\quad i\in [1,k-1], \; s\in [0,k-1],
$$
Then we have
$$
\d_{n^i}+\d_{n}^{i+j}+\cdots+\d_{n}^{i+(k-1)j}\in \Span({\cal B}_{{\cal V}_n}).\\
$$
That is,
$$
\Omega^{n-1}\bigcap  \Span({\cal B}_{{\cal V}_n})\neq \{0\},
$$
which implies that (\ref{5.1.5}) is a linearly dependent set.

\hfill $\Box$

From the proof of Theorem \ref{t5.1.1}, one sees easily how to choose elements of ${\cal B}_{{\cal V}_n}$ to form ${\cal D}_{{\cal V}_n}$, and hence the basis of $\Omega^n$, which is ${\cal D}^n_{{\cal V}}=\bigcup_{1\leq s\leq n}  {\cal D}_{{\cal V}_s}$.

\begin{cor}\label{c5.1.2}
\begin{itemize}
\item[(i)] Assume $n$ has no multi-divisor, then
\begin{align}\label{5.1.10}
{\cal D}_{{\cal V}_n}={\cal B}_{{\cal V}_n}.
\end{align}
\item[(ii)]  Assume $n$ has multi-divisor. Express $n=s^2\cdot j$, where $j$ has no multi-divisor. (Such an expression is unique.)
Then
\begin{align}\label{5.1.11}
{\cal D}_{{\cal V}_n}=\left\{ \d_n^j\in {\cal B}_{{\cal V}_n}\;|\; j\leq (s-1)s\cdot j\right\} 
\end{align}
\end{itemize}
\end{cor}

\begin{exa}\label{e5.1.3}
\begin{itemize}
\item[(i)] Consider $n=12:= s^2\cdot j$. Then $s=2$ and $j=3$, and
$$
{\cal B}_{{\cal V}_n}=\{\d_{12}^1,\d_{12}^5,\d_{12}^7,\d_{12}^{11}\}.
$$
Now we calculate $(s-1)s\cdot j=6$. Hence,
$$
{\cal D}_{{\cal V}_n}=\{\d_{12}^1,\d_{12}^5\}.
$$

Note that
$$
\d_{12}^1+\d_{12}^7=\J_2\otimes \d_6^1,
$$
and
$$
\d_{12}^5+\d_{12}^{11}=\J_2\otimes \d_6^5,
$$
Hence, remove $\d_{12}^5$ and $\d_{12}^{11}$ is reasonable.

\item[(ii)] Consider $n=27:= s^2\cdot j$. Then $s=3$ and $j=3$, and
$$
\begin{array}{l}
{\cal B}_{{\cal V}_n}=
\{\d_{27}^1,\d_{27}^2,\d_{27}^4,\d_{27}^{5},
 \d_{27}^{7},\d_{27}^{8},\d_{27}^{10},\d_{27}^{11},\\
 \d_{27}^{13},\d_{27}^{14},\d_{27}^{16},\d_{27}^{17},
  \d_{27}^{19},\d_{27}^{20},\d_{27}^{22},\d_{27}^{23},
  \d_{27}^{25},\d_{27}^{26}\}.\\
\end{array}
$$
Calculate $(s-1)s\cdot j=18$. Hence,
$$
\begin{array}{l}
{\cal D}_{{\cal V}_n}=
\{\d_{27}^1,\d_{27}^2,\d_{27}^4,\d_{27}^{5},
 \d_{27}^{7},\d_{27}^{8},\d_{27}^{10},\d_{27}^{11},\\
 \d_{27}^{13},\d_{27}^{14},\d_{27}^{16},\d_{27}^{17}\}.\\
\end{array}
$$
To see the rest elements have to be deleted, we have
$$
\J_3\otimes \d_9^k=\d_{27}^k+\d_{27}^{k+9}+\d_{27}^{k+18},\quad k\in [1,8].
$$
As $k$ goes from $1$ to $8$, one sees easily that
$\d_{27}^{19},\d_{27}^{20},\cdots,\d_{27}^{26}$ have to be removed.
\end{itemize}
\end{exa}

Now we are able to construct a basis of $\Omega$. Using Corollary \ref{c5.1.2}, the basis elements can be obtained. A few leading elements are listed as follows.

\begin{align}\label{5.1.12}
\begin{array}{ccccc}
d_1=1,&d_2=\d_2^1,&d_3=\d_3^1,&d_4=\d_3^2,&d_5=\d_4^1,\\
d_6=\d_5^1,&d_7=\d_5^2,&d_8=\d_5^3,&d_9=\d_5^4,&d_{10}=\d_6^1,\\
d_{11}=\d_6^5,&d_{12}=\d_7^1,&d_{13}=\d_7^2,&d_{14}=\d_7^3,&d_{15}=\d_7^4,\\
d_{16}=\d_7^5,&d_{17}=\d_7^6,&d_{18}=\d_8^1,&d_{19}=\d_8^3,&d_{20}=\d_9^1,\\
d_{21}=\d_9^2,&d_{22}=\d_9^4,&d_{23}=\d_9^5,&d_{24}=\d_{10}^1,&d_{25}=\d_{10}^3,\\
d_{26}=\d_{10}^7,&d_{27}=\d_{10}^9,&\cdots&~&~\\
\end{array}
\end{align}

Since $\Omega$ is an inner product, using Gram-Schmidt orthonormalization algorithm, we can get orthonormal basis as
\begin{align}\label{5.1.13}
\begin{array}{l}
e_1=1,~~ e_2=(1,-1)^T,\\
e_3=\sqrt{1/2}(2,-1,-1)^T,~~ e_4=\sqrt{3/2}(0,1,-1)^T,\\
e_{5}=\sqrt{2}(1,0,-1,0)^T,~~e_{6}=\frac{1}{2}(4,-1,-1,-1,-1)^T,\\
\e_{7}=\sqrt{5/12}(0,3,-1,-1,-1)^T,\\\e_8=\sqrt{5/6}(0,0,2,-1,-1)^T\\
\e_{9}=\sqrt{5/2}(0,0,0,1,-1)^T,~~\cdots\\
\end{array}
\end{align}

\begin{rem}\label{r5.1.4}
\begin{itemize}
\item[(i)] The orthonormal basis can easily be obtained up to finite terms via computer numerically.
\item[(ii)] Using orthonormal basis, $\Omega$ can be imbedded isometrically into Hilbert space $\ell_2$ \cite{xue23}.
\item[(iii)] Inspired that the discussion of this subsection is about $\Omega=\Omega_{\ell}$, since the elements in the basis are neither left reducible no right reducible, it is easy to see that the basis of $\Omega_{\ell}$ is also the basis of $\Omega_{r}$.
\end{itemize}
\end{rem}

\subsection{Norms of Signal Space}

Consider $x=(x_1,x_2,\cdots)$, which is a sequence of real numbers, denote it by $E^{\infty}$, which is a vector space.  We can define a set of norms on $E^{\infty}$ as
\begin{itemize}
\item $\ell_0$ norm:
\begin{align}\label{5.2.1}
\|x\|_0:=w(x).
\end{align}
Then
\begin{align}\label{5.2.2}
\ell_0:=\{x\in E^{\infty}\;|\;\|x\|_0<\infty\}.
\end{align}
\item $\ell_p$ norm:
\begin{align}\label{5.2.3}
\|x\|_p:=\left(\dsum_{i=1}^{\infty} |x_i|^p\right)^{1/p},\quad p=1,2,\cdots.
\end{align}
Then
\begin{align}\label{5.2.4}
\ell_p:=\{x\in E^{\infty}\;|\;\|x\|_p<\infty\},
quad p=1,2,\cdots.
\end{align}

\end{itemize}

Then the following facts are well known \cite{tay80}.

\begin{prp}\label{p5.2.1}
\begin{itemize}
\item[(i)] $\ell_0$ is a Fr\`{e}chet space \footnote{On a vector space $V$,  a mapping $\|\cdot\|$ is called a pseudo-norm, if
\begin{itemize}
\item
$$
\|x\|\geq 0, ~\mbox{and}~\|x\|=0 ~\mbox{implies}~ x=0.
$$
\item
$$
\|x+y\|\leq \|x\|+\|y\|,\quad x,y\in V,
$$
\item
$$
\|-x\|=\|x\|.
$$
\end{itemize}
A complete pseudo-normed space is called a Fr\`{e}chet space.}

\item[(ii)] $\ell_p$, $p\geq 1$ are Banach spaces.

\item[(iii)] $\ell_2$ is an Hilbert space.
\end{itemize}

\end{prp}

Now go back to the set of signals. If a signal $x\in \R^n$. Then it is easy to be embedded into $E^{\infty}$ as
$$
\psi: x\mapsto (x_1,\cdots,x_n,0,0\cdots).
$$
In this way, $x\in \ell_p$, $\forall p\geq 0$. This $\psi$ is commonly used in signal processing community. But if we consider infinite-dimensional signal or the infinite union of finite dimensional signals \cite{dua11}, we must be very careful.

Note that $E^{\infty}$ is a vector space, but it is not a topological space. All $\ell_p$ are its subspaces. Using the distance deduced by the norm (pseudo-norm), all $\ell_p$ become topological spaces. The topologies deduced by the distances are denoted by ${\cal T}_{\ell_p}$.

Consider  $\R^{\infty}$. As aforementioned on $\R^{\infty}$ there are two commonly used topologies: natural topology  (${\cal T}_n$) and distance deduced topology (${cal T}_d$).

Consider the natural topology first. Under this topology, any two points $p,q\in \R^{\infty}$, $p\neq q$, are separable (under $T_2$, or Hausdorff sense).
Hence,
\begin{align}\label{sp.3.2.5}
\psi\left(\R^{\infty}\right)\subset \ell_p,\quad \forall p\geq 0.
\end{align}
Because each element in $\R^{\infty}$ has only finite terms.

The mapping   $\psi$, defined by (\ref{sp.3.2.5}) is one-to-one, hence $\psi$  can be considered as an imbedding. And then we can pose the subspace topology of $\ell_p$  to  $\R^{\infty}$. For instance, let $p=2$. Then $\R^{\infty}$ is an inner product space. But it is not a Hilbert space, because it is not complete.

All  spaces
$$
(\R^{\infty},{\cal T}_{\ell_p}),\quad p \geq 0,
$$
are topological spaces. Some properties follow from the definition immediately.

\begin{prp}\label{psp.3.2.2}
\begin{itemize}
 \item[(i)] $(\R^{\infty},{\cal T}_{\ell_p})$ homeomorphic to neither $(\R^{\infty},{\cal T}_n)$ no $(\R^{\infty},{\cal T}_d)$.

 \item[(ii)]   If $p\neq q$, then $(\R^{\infty},{\cal T}_{\ell_p})$ does not homeomorphic to $(\R^{\infty},{\cal T}_{\ell_q})$.

 \item[(iv)] Consider $R^n$ , which is considered as the space of signals with fixed length. Assume it has the subspace topology of  $\R^{\infty}$,  then
 \begin{align}\label{sp.3.2.6}
  (\R^{n},{\cal T}_{\ell_p}|_{\R^n})\cong (\R^{n},{\cal T}_{n}|_{\R^n})\cong (\R^{n},{\cal T}_{d}|_{\R^n}).
\end{align}
\end{itemize}
\end{prp}

\begin{rem}\label{rsp.3.2.3}
(\ref{sp.3.2.6}) shows that when the signals have a fixed dimension, all the vector space topologies are equivalent. Only the dimension varying signals or infinity dimensional signals are considered, the different topologies become meaningful.
\end{rem}

Next, we consider the distance deduced topology (${cal T}_d$).

Assume $\{e_i\;|\;i=1,2,\cdots\}$ is an orthonormal basis of $\Omega=\R^{\infty}/\lra$ .
Consider $x=(x_1,\cdots,x_n)^T\in \R^n$. Then
$$
\bar{x}=\dsum_{i=1}^{m_n}\xi_ie_i,
$$
where,
$$
m_n=|{\cal D}^n_{{\cal V}}| \leq n.
$$
Define $\phi:\R^{\infty}\ra E^{\infty}$ as
\begin{align}\label{sp.3.2.7}
\phi(x):=(\xi_1,\xi_2,\cdots,\xi_{m_n},0,0,\cdots)
\end{align}

Now consider $\R^{\infty}$. It is obvious that
\begin{align}\label{5.2.5}
\psi\left(\R^{\infty}\right)\subset \ell_p,\quad \forall p\geq 0.
\end{align}
Because each element is $\R^{\infty}$ has only finite terms.

Since the $\psi$ in (\ref{5.2.5}) is one=to-one, then $\psi$ can be considered as an embedding.  Hence, we can pose $\R^{\infty}$ the subspace topology of $\ell_p$. For instance, we assume $p=2$. Then $\R^{\infty}$ becomes an inner product space. Note that
$$
(\R^{\infty},{\cal T}_{\ell_p}),\quad p \geq 0,
$$
are all topological spaces. According to the definitions, the following proposition is easily verifiable.

\begin{prp}\label{p5.2.2}
\begin{itemize}
 \item[(i)] $(\R^{\infty},{\cal T}_{\ell_p})$ is homeomorphic to neither $(\R^{\infty},{\cal T}_n)$ no $(\R^{\infty},{\cal T}_d)$.

 \item[(ii)]   If $p\neq q$, then $(\R^{\infty},{\cal T}_{\ell_p})$ is not homeomorphic to $(\R^{\infty},{\cal T}_{\ell_q})$.

 \item[(iv)] Consider $R^n$ with the inherited (subspace) topology of $\R^{\infty}$. Then
 \begin{align}\label{5.2.6}
  (\R^{n},{\cal T}_{\ell_p}|_{\R^n})\cong (\R^{n},{\cal T}_{n}|_{\R^n})\cong (\R^{n},{\cal T}_{d}|_{\R^n}).
\end{align}
\end{itemize}
\end{prp}

\begin{rem}\label{r5.2.3}
(\ref{5.2.6}) shows that when signals of fixed dimension are considered, any topologies are the same. Only when the dimension-varying signals are investigated, the different topologies may cause different results.
\end{rem}

\section{Dimension-free STP-CS}

According to Corollary \ref{c3.3.3}, if all signals are expressed as elements in signal space,  the left STP-CS can be formally expressed
\begin{align}\label{4.1.1}
\bar{y}=\A_{\ell} \lvtimes \bar{x},
\end{align}
where $\bar{x}\in \Omega_{\ell}$ is the original signal, $\bar{y}\in \Omega_{\ell}$ is the sampled data, and $\A_{\ell}\in \Sigma_{\ell}$ is the sensing matrix.

We call (\ref{4.1.1}) a dimension-free STP-CS, because it can be used to treat signals of arbitrary dimensions.

Correspondingly, if we take $\Omega_r$ as the signal space, then the right STP-CS can be expressed as
\begin{align}\label{4.1.2}
\bar{y}=\A_{r} \lvtimes \bar{x},
\end{align}
where $\bar{x}\in \Omega_{r}$ is the original signal, $\bar{y}\in \Omega_{r}$ is the sampled data, and $\A_{r}\in \Sigma_{r}$ is the sensing matrix.

Both the expressions (\ref{4.1.1}) and (\ref{4.1.2}) are dimension-free. That is, there is no dimension restriction on the original or compressed signal. But in practical use, dimension depending expression is more convenient.
Let us consider a particular (matrix-vector) form for STP-CS: We can w.l.g., assume that $A$ is left irreducible. Otherwise, we can use irreducible $A_0\in \A$ to replace $A$.

\begin{prp}\label{t4.1.1}
\begin{itemize}
\item[(i)] Assume $x\in \R^p$, $A\in {\cal M}_{m\times n}$, and $p=sn$. Then (\ref{4.1.1}) becomes
\begin{align}\label{4.1.3}
y=(A\otimes I_s)x,
\end{align}
which is essentially the same as (\ref{1.2}).

\item[(ii)] Assume $x\in \R^p$, $A\in {\cal M}_{m\times n}$, and $\lcm(n,p)=t$, where $t=sn=rp$, $r>1$. Then (\ref{4.1.1}) becomes
\begin{align}\label{4.1.4}
y=(A\otimes I_s)(x\otimes \J_r).
\end{align}

\item[(iii)] Dimension-varying signal:
Assume $D=\{d_i\:\:i\in [1,\ell]\}\subset \Z^+\}$ is a finite set, and $x\in \R^p$, $p\in D$. Then there exists a switching signal
$\sigma(t)\in D$, $t\geq 0$, such that
\begin{align}\label{4.1.5}
y(t)=(A\otimes I_{s(\sigma(t))})(x(t)\otimes \J_{r(\sigma(t))}).
\end{align}
\end{itemize}
\end{prp}

Correspondingly, using right system, we have

\begin{prp}\label{t4.1.2}
\begin{itemize}
\item[(i)] Assume $x\in \R^p$, $A\in {\cal M}_{m\times n}$, and $p=sn$. Then (\ref{4.1.2}) becomes
\begin{align}\label{4.1.6}
y=(I_s\otimes A)x.
\end{align}

\item[(ii)] Assume $x\in \R^p$, $A\in {\cal M}_{m\times n}$, and $\lcm(n,p)=t$, where $t=sn=rp$, $r>1$. Then (\ref{4.1.2}) becomes
\begin{align}\label{4.1.7}
y=(I_s\otimes A)(\J_r\otimes x).
\end{align}

\item[(iii)] Dimension-varying signal:
Assume $D=\{d_i\:\:i\in [1,\ell]\}\subset \Z^+\}$ is a finite set, and $x\in \R^p$, $p\in D$. Then there exists a switching signal
$\sigma(t)\in D$, $t\geq 0$, such that
\begin{align}\label{4.1.8}
y(t)=(I_{s(\sigma(t))}\otimes A)(\J_{r(\sigma(t))}\otimes x(t)).
\end{align}
\end{itemize}
\end{prp}

In practical use, the conventional matrix expression is necessary.

\begin{itemize}
\item[(i)] Case 1:

Assume $n|p$ with $p=ns$.
\end{itemize}
This case is commonly assumed in current literature.

We consider the right STP-CS first. Then we have (\ref{4.1.6}).

\begin{prp}\label{p4.1.3} Consider (\ref{4.1.6}). Then
\begin{align}\label{4.1.9}
\spark(I_s\otimes A)=\spark(A).
\end{align}
\end{prp}

\noindent{\it Proof.} Since
\begin{align}\label{4.1.10}
I_s\otimes A=\begin{bmatrix}
A&0&\cdots&0\\
0&A&\cdots&0\\
\vdots&~&~&~\\
0&0&\cdots&A\\
\end{bmatrix}.
\end{align}
Then it is clear that the smallest set of linearly dependent columns can only be found within each $A$. This fact leads to (\ref{4.1.9}).

\hfill $\Box$

Using Proposition \ref{p1.1},  we have the following result.

\begin{cor}\label{c4.1.4} Consider (\ref{4.1.6}). If $\spark(A)>2k$, then for each $y\in \R^{sm}$ there is at most one solution $x\in \Sigma^p_k$.
\end{cor}

In fact, we can get a better result for STP-CS.

Let $x=(x^1,x^2,\cdots,x^s)^T$, where $x^i\in \R^n$, $i\in [1,s]$. Define
\begin{align}\label{4.1.11}
\Sigma^p_{k/n}:=\{x=(x^1,x^2,\cdots,x^s)^T\in \R^p\;|\; \forall x^i\in \Sigma^n_k\}.
\end{align}

\begin{prp}\label{p4.1.5} Consider (\ref{4.1.6}). If $\spark(A)>2k$, then for each $y\in \R^{sm}$ there is at most one solution $x\in \Sigma^p_{k/n}$.
\end{prp}

\noindent{\it Proof.} Using (\ref{4.1.10}) we can get the following decomposed system as
\begin{align}\label{4.1.12}
\begin{cases}
y^1=Ax^1,\\
y^2=Ax^2,\\
\vdots,\\
y^s=Ax^2.
\end{cases}
\end{align}
Using  Proposition \ref{p1.1} to each sub-equations, the conclusion follows.

\hfill $\Box$

Next, we consider left STP-CS.  Then we have (\ref{4.1.3}).

\begin{lem}\label{l4.1.6} \cite{che12} Let $A\in {\cal M}_{m\times n}$, $B\in {\cal M}_{p\times q}$. Then
\begin{align}\label{4.1.13}
W_{[m,p]}(A\otimes B)W_{q,n}=B\otimes A.
\end{align}
\end{lem}

Define
$$
z(x)=W_{[n,s]} x:=(z^1,z^2,\cdots,z^s)^T.
$$
It is well known that $z$ is obtained from $x$ by an element permutation. Then
similarly to Proposition \ref{p4.1.5}, we have the following result.

\begin{prp}\label{p4.1.7} Consider (\ref{4.1.3}).
If $\spark(A)>2k$, then for each $y\in \R^{sm}$ there is at most one solution $x$ with $z(x)\in \Sigma^p_{k/n}$.
\end{prp}

\noindent{\it Proof.}
Using Lemma \ref{l4.1.6}, we have
\begin{align}\label{4.1.14}
\begin{array}{l}
(A\otimes I_s)x\\
=W_{[s,m]}(I_s\otimes A)W_{[n,s]}x\\
=W_{[s,m]}(I_s\otimes A)z\\
=y.
\end{array}
\end{align}
\eqref{4.1.14} is equivalent to
\begin{align}\label{4.1.15}
(I_s\otimes A)z=W^{-1}_{[s,m]}y=W_{[m,s]}y.
\end{align}
The conclusion follows from Proposition \ref{p4.1.5} immediately.

\hfill $\Box$

\begin{itemize}
\item[(ii)] Case 2:
Assume $n\nmid p$ with $\lcm(n,p)=t$.
\end{itemize}

Then the (\ref{4.1.1}) becomes
\begin{align}\label{4.1.16}
(A\otimes I_{t/n})\tilde{x}=\tilde{y},
\end{align}
where $\tilde{x}=x\otimes \J_{t/p}$, $\tilde{y}=y\otimes \J_{tm/n}$. (\ref{4.1.16}) has exactly the same form as (\ref{4.1.6}).

the (\ref{4.1.2}) becomes
\begin{align}\label{4.1.17}
(I_{t/n}\otimes A)\tilde{x}=\tilde{y},
\end{align}
where $\tilde{x}= \J_{t/p}\otimes x$, $\tilde{y}= \J_{tm/n}\otimes y$. (\ref{4.1.17}) has exactly the same form as (\ref{4.1.3}).

We conclude that Case 2 can be converted into Case 1. Then the results for Case 1 are available for Case 2.

\begin{dfn}\label{d4.1.8} Let $\A\in \Omega_{\ell}$ (or $\A\in \Omega_{r}$ ). Then
\begin{align}\label{4.1.18}
\spark(\A_{\ell}):=\spark(A_0),\quad (\mbox{or}~\spark(\A_{r}):=\spark(A_0),
\end{align}
where $A_0\in \A_{\ell}$ is left  irreducible ($A_0\in \A_{r}$ is right  irreducible) .
\end{dfn}

Next, we consider some further properties.

\begin{dfn}\label{d4.1.9}   (Coherence)

Assume $A=A_0\otimes I_s$, where $A_0$ is left irreducible (or $A=I_t\otimes A_0$, where $A_0$ is right irreducible). Then the coherence of $\A$ is defined as
\begin{align}\label{4.1.19}
\mu(\A):=\mu(A_0).
\end{align}
\end{dfn}

Finally, we consider the RIP. Denote by
\begin{align}\label{4.1.20}
\Sigma_{k/n}=\bigcup_{p\geq n}\Sigma^p_{k/n}.
\end{align}
Then we can define dimension-free RIP as follows.

\begin{dfn}\label{d4.1.10}

$\A$ with $A_0\in {\cal M}_{m\times n}$ is said to satisfy the $(k,\d)$-RIP, if there exists  $\d=\d_k^A\in (0,1)$, such that
\begin{align}\label{4.1.21}
(1-\d)\|x\|_{{\cal V}}^2\leq \|A_0\rvtimes x\|_{{\cal V}}^2\leq (1+\d)\|x\|_{{\cal V}}^2,\quad \forall x\in \Sigma_{k/n}.
\end{align}
\end{dfn}

\begin{rem}\label{r4.1.11}
\begin{itemize}
\item[(i)] It is easy to verify that for any $A\in \A$, the coherence of $A$ is the same. Hence
the Definition \ref{d4.1.9} 
is reasonable. In fact, $A_0$ can be replaced by any $A\in \A$.

\item[(ii)] The inner product and norm can also be replaced by $\langle \cdot, \cdot \rangle_{{\cal V}}$ and $\|\cdot \|_{{\cal V}}$. which are related to the topological structure of $\Omega$.

\item[(iii)] Definition  \ref{d4.1.9} 
seems more restrictive than the original definition for fixed dimensional case. In fact, it is not. Note that if $A\sim B$ and $x\lra y$, then
    $$
    \|A\ltimes x\|_{{\cal V}}=\|B\ltimes y\|_{{\cal V}}.
    $$
 Hence if $A_0$ satisfies the $(k,\d)$-RIP as defined in Definition \ref{d1.101}, then $\A$ satisfies   the $(k,\d)$-RIP as defined in Definition \ref{d4.1.10}.
 Hence, the fixed dimension $(k,\d)$-RIP implies the dimension-free $(k,\d)$-RIP, which ensures the precise reconstruction for much more signals.

\end{itemize}
\end{rem}

\section{BIBD-Based Sensing Matrix}

\subsection{BIBD-Based Construction}

This subsection gives a briefly review for the construction of sensing matrix based on  balanced incomplete block design (BIBD), proposed in \cite{lia22}.

\begin{dfn}\label{d4.2.1} Consider a matrix $A\in {\cal M}_{\a\times \b}$.
\begin{itemize}
\item[(i)]  $A$ is called a sign matrix  if
$$
a_{i,j}\in \{1,-1\},\forall i\in [1,\a], j\in [1,\b].
$$
\item[(ii)] $A$ is called a Boolean matrix if
$$
a_{i,j}\in \{1,0\},\forall i\in [1,\a], j\in [1,\b].
$$
The column degree of $A$ is denoted by
$$
d_c(\Col_j(A))=\dsum_{i=1}^{\a}a_{i,j},\quad j\in [1,\b].
$$
The row degree of $A$ is denoted by
$$
d_r(\Row_i(A))=\dsum_{j=1}^{\b}a_{i,j},\quad i\in [1,\a].
$$
\end{itemize}
\end{dfn}

\begin{dfn}\label{d4.2.2} \cite{lia22} Let $X=\{x_1,x_2,\cdots,x_{\a}\}$ and $\mathbb{P}=\{P_1,P_2,\cdots,P_{\b}\}\subset 2^X$, i.e., each block $P_j\in 2^X$, $j\in [1,\b]$.
\begin{itemize}
\item[(i)] $H\in {\cal B}_{\a\times \b}$ is defined by
$$
h_{i,j}=\begin{cases}
1,\quad x_i\in P_j,\\
0,\quad Otherwise.
\end{cases}
$$
$H$ is called an incidence matrix.

\item[(ii)] $\mathbb{P}$ is said to have BIBD, if its index matrix satisfies the following conditions.
\begin{itemize}
\item Each element $x_i$ appears exactly in $r$ blocks.
\item Each block contains exactly $2\leq k<\a$ elements, i.e., $|P_j|=k$, $\forall j$.
\item Each pair of elements in X appears exactly in $\lambda$ blocks.
\end{itemize}
Precisely, $\mathbb{P}$ is said to have a $(\a,\b,r,k,\lambda)$-BIBD.

\end{itemize}

\end{dfn}

Assume $X=\{x_1,x_2,\cdots,x_{\a}\}$ and $\mathbb{P}=\{P_1,P_2,\cdots,P_{\a}\}$. \cite{lia22} proposed a way to construct a deterministic sensing matrix as follows.

Assume $\mathbb{P}$ has $(\a,\a, \a-1,\a-1,\a-
2)$-BIBD, then its incidence matrix can be expressed as
\begin{align}\label{4.2.1}
H=\begin{bmatrix}
1&1&\cdots&1&0\\
1&1&\cdots&0&1\\
\vdots&~&~&~&~\\
1&0&\cdots&1&1\\
0&1&\cdots&1&1\\
\end{bmatrix}
\end{align}

The sensing matrix $A$ is constructed through two expansions.

\begin{itemize}
\item Vertical Expansion:
\end{itemize}

\begin{alg}\label{a4.2.3} Assume an incidence matrix $H$ is given.
\begin{itemize}
\item Step 1: Keep first column unchanged. Start from second column. Keep the first $1$ in column 2 unchanged. If the second $1$ meets $0$ at the same row of the first column, keep this $1$ unchanged. Otherwise,  move all remaining elements (including this $1$) down, till its same row element in first column being $0$. Then do the same thing for third $1$ and keep doing for all other $1$ elements one by one in the order.
\item Step 2: For third column. Allow it has one $1$, which is on the same row of the $1$ for first and second columns. Otherwise, move this $1$ and all the below elements down. Until the above requirement satisfies, which keeps the inner product of the first or second column with third one is $1$.
\item Step k-1: For $k$-th column,  similarly to third column, as long as the inner product of $k$-th column with one of the first $k-1$ columns is greater than $1$, move the corresponding $1$ with all below elements down, until the inner product requirement is satisfied.
\end{itemize}
\end{alg}

We give a simple example to depict this algorithm.
\begin{exa}\label{e4.2.4} \cite{lia22} Assume $\a=4$, the incidence matrix is
$$
H=\begin{bmatrix}
1&1&1&0\\
1&1&0&1\\
1&0&1&1\\
0&1&1&1\\
\end{bmatrix}
$$
Its vertical expanded matrix becomes
$$
H_v=\begin{bmatrix}
1&1&1&0\\
1&0&0&1\\
1&0&0&0\\
0&1&0&1\\
0&1&0&0\\
0&0&1&1\\
0&0&1&0\\
\end{bmatrix}
$$
\end{exa}

It was proved in  \cite{lia22} that $H_v\in {\cal M}_{m\times \a}$, where
\begin{align}\label{4.2.2}
m=\a^2-3\a+3.
\end{align}
Moreover, the coherence of $H_v$ is
\begin{align}\label{4.2.3}
\mu(H_v)=\frac{1}{\a-1}.
\end{align}

\begin{itemize}
\item Horizontal Expansion:
\end{itemize}

\begin{dfn}\label{d4.2.5}    Let $A\in {\cal M}_{(\a-1)\times r}$  be a sign matrix, and $D\in {\cal M}_{r\times r}$ be a nonsingular diagonal matrix, say $D=\diag(d_1,d_2, \cdots, d_r)$, where $d_s >0$, $s\in [1,r]$ and $d_i\neq d_j$, $i\neq j$.
Then $B:=AD$ is called an embedding matrix.
\end{dfn}

\begin{alg}\label{a4.2.6} Assume the vertically expanded incident matrix $H_v$ and an embedding matrix $B$ are given.

For each column of  $H_v$, say, $\Col_j(H_v)$, its first $1$ is replaced by $\Row_1(B)$, second $1$ is replaced by $\Row_2(B)$, $\cdots$, till last $1$ is replaced by $\Row_{\a-1} (B)$. $0$ is replaced by $\underbrace{0,0,\cdots,0}_r$.
The resulting matrix is expressed as
\begin{align}\label{4.2.4}
A:=H_v\odot B.
\end{align}
\end{alg}

The following result and the estimation (\ref{1.6}) are fundamental for this design.

\begin{thm}\label{t4.2.7} \cite{ami12} Let $H\in {\cal M}_{\a\times \b}$ be an incidence matrix with column degree
$$
d_c(\Col_j(H))=d,\quad j\in [1,\b],
$$
$B\in {\cal M}_{d\times s}$ be an embedding matrix, and $\Phi=H\odot B$. Then
\begin{align}\label{4.2.5}
\mu(\Phi)=\max\{\mu(H),\mu(B)\}.
\end{align}
\end{thm}\footnote{The original requirement for $B$ is \cite{ami12} ``the elements of $B$ have the same absolute values in the same column, but the elements have different absolute values in different columns." Such matrices can be constructed as in the Definition \ref{4.2.5} for embedding matrix.}

\begin{cor}\label{c4.2.8} Using BIBD-based design, the best solution is
\begin{align}\label{4.2.6}
\mu(\Phi)=\mu(H_v)=\frac{1}{\a-1},
\end{align}
which can be reached as long as
\begin{align}\label{4.2.7}
\mu(B)\leq \frac{1}{\a-1}.
\end{align}
\end{cor}

\begin{exa}\label{e4.2.9}  Recall Example \ref{e4.2.4}. It is easy to find an embedding matrix as
\begin{align}\label{4.2.8}
B=\begin{bmatrix}
1&1&1&-1\\
1&-1&1&1\\
1&1&-1&1
\end{bmatrix}
\end{align}.
Then the sensing matrix can be constructed as
\begin{align}\label{4.2.8}
\begin{array}{l}
\Phi:=H_v\odot B\\
=\left[
\begin{array}{cccccccc}
1&1&1&-1&1&1&1&-1\\
1&-1&1&1&0&0&0&0\\
1&1&-1&1&0&0&0&0\\
0&0&0&0&1&-1&1&1\\
0&0&0&0&1&1&-1&1\\
0&0&0&0&0&0&0&0\\
0&0&0&0&0&0&0&0\\
\end{array}\right.\\
\left.
\begin{array}{cccccccc}
1&1&1&-1&0&0&0&0\\
0&0&0&0&1&1&1&-1\\
0&0&0&0&0&0&0&0\\
0&0&0&0&1&-1&1&1\\
0&0&0&0&0&0&0&0\\
1&-1&1&1&1&1&-1&1\\
1&1&-1&1&0&0&0&0\\
\end{array}\right].\\
\end{array}
\end{align}.

It is easy to verify that
$$
\mu(B)=\frac{1}{\a-1}=\frac{1}{3}.
$$
Hence,
$$
\mu(\Phi)=\mu(H_v)=\frac{1}{3}.
$$
\end{exa}

\cite{ton21} provides some results for constructing matrix $B$, which meets (\ref{4.2.7}). Some other examples are Hadamard matrix or DFT matrix and certain their generalizations.  Unfortunately, the latter  are not embedding matrix. We will look for embedding matrices which meets (\ref{4.2.7}).

\subsection{A modified BIBD-Based Sensing Matrix}

This subsection aims on improving the BIBD-based sensing matrix.

First, we improve the vertical expansion.

\begin{dfn}\label{d4.3.1} Let $H\in {\cal M}_{\a\times \a}$  be an incident matrix with column degree $\a-1$. A new vertical expansion, denoted by $H_*$, is defined as follows.
\begin{align}\label{4.3.1}
H_*=\begin{bmatrix}
H_1\\
H_2\\
\vdots\\
H_{\a-1}
\end{bmatrix}\in {\cal M}_{m\times \a},
\end{align}
where
$$
\begin{array}{l}
H_1=\left[\J_{\a-1}, I_{\a-1}\right],\\
H_2=\left[0_{\a-2},\J_{\a-2}, I_{\a-2}\right],\\
\vdots\\
H_{\a-2}=\left[\underbrace{0_2,\cdots,0_2}_{\a-3}, \J_2,I_2\right],\\
H_{\a-1}=\left[\underbrace{0,\cdots,0}_{\a-2},1,1\right].
\end{array}
$$
\end{dfn}

\begin{exa}\label{e4.3.2} Recall the $H$ in Example \ref{e4.2.4}. Then
$$
H_*=
\begin{bmatrix}
1&1&0&0\\
1&0&1&0\\
1&0&0&1\\
0&1&1&0\\
0&1&0&1\\
0&0&1&1
\end{bmatrix}
$$
\end{exa}

The following proposition comes from the construction immediately.

\begin{prp}\label{p4.3.3}  Let $H\in {\cal M}_{\a\times \a}$  be an incident matrix with column degree $\a-1$. Then the vertically expanded matrix $H_*$ satisfies the following conditions.
\begin{itemize}
\item[(i)] $H_*\in {\cal M}_{m\times \a}$, where
\begin{align}\label{4.3.101}
m=\frac{\a(\a-1)}{2}.
\end{align}
\item[(ii)]
\begin{align}\label{4.3.102}
\begin{array}{l}
 d_c(\Col_j(H_*)=\a-1,\quad j\in [1,\a];\\
 d_r(\Row_i(H_*)=2,\quad i\in [1,m].\\
 \end{array}
 \end{align}
\item[(iii)] The coherence
$$
\mu(H_*)=\frac{1}{\a-1}.
$$
\end{itemize}
\end{prp}

Comparing with $H_v$, it is obvious that $H_*$ has the same column degree and the same coherence with $H_v$. Moreover,
$H_*$ has less rows ($\frac{\a(\a-1)}{2}$) comparing with $H_v$ ($\a^2-3\a+3$). The smaller the $m$ is, the higher the compress rate is obtained.

Next, we consider how to construct an embedding matrix to meet (\ref{4.2.7}).

\begin{lem}\label{l4.3.4}  Assume $B=AD$ is an embedding matrix, where $A\in {\cal M}_{\a\times \b}$  is a sign matrix, and $D=\diag(d_1,\cdots,d_{\b})$ is a nonsingular diagonal matrix with $d_i>0$, $i\in [1,\b]$ and $d_i\neq d_j$, $i\neq j$. Then
\begin{align}\label{4.3.2}
\mu(B)=\mu(A).
\end{align}
\end{lem}

\noindent{\it Proof.} Denote $A_i:=\Col_i(A)$.
$$
\begin{array}{ccl}
\mu(B_i,B_j)&=&\frac{\langle d_iA_i,d_jA_j\rangle}{\|d_iA_i\|\|d_jA_j|}\\
~&=&\frac{|d_i||d_j|\langle A_i,A_j\rangle}{|d_i||d_j|\|A_i\|\|A_j|}\\
~&=&\mu(A_i,A_j),\quad i\neq j.
\end{array}
$$
The conclusion follows.
\hfill $\Box$

It is clear now to construct an embedding matrix, we need only to construct a sign matrix, satisfied (\ref{4.2.7}).

\begin{prp}\label{p4.3.5} Let $A\in {\cal M}_{\a\times \b}$ be a sign matrix.
\begin{itemize}
\item[(i)] Assume $\a$ is an even number. Then $A$ satisfies  (\ref{4.2.7}), if and only if,
\begin{align}\label{4.3.3}
\langle A_i,A_j\rangle =0,\quad i\neq j.
\end{align}
\item[(ii)] Assume $\a$ is an odd number. Then $A$ satisfies  (\ref{4.2.7}), if and only if,
\begin{align}\label{4.3.4}
\langle A_i,A_j\rangle =\pm 1,\quad i\neq j.
\end{align}
\end{itemize}
\end{prp}

\noindent{\it Proof.}
\begin{itemize}
\item[(i)] Assume $t$ is even. Set
$$
\begin{array}{l}
N_+=|\{k\;|\;a_{k,i}a_{k,j}=1\}|,\\
N_-=|\{k\;|\;a_{k,i}a_{k,j}=-1\}|.
\end{array}
$$
Then
$$
N_+-N_-=0,\pm 2, \pm 4,\cdots.
$$
Then the corresponding coherence
$$
\mu(A_i ,A_j)=0,2/\a,4/\a,\cdots.
$$
Hence only when $N_+=N_-$,   (\ref{4.2.7}) is satisfied, which leads to (\ref{4.3.3}).
\item[(ii)] Assume $\a$ is odd. Then
$$
N_+-N_-=\pm 1,\pm 3, \pm 5,\cdots.
$$
The corresponding coherence
$$
\mu(A_i ,A_j)=1/\a,3/\a,5/\a,\cdots.
$$
Hence only when $N_+N_-=\pm1$,   (\ref{4.2.7}) is satisfied, which leads to (\ref{4.3.4}).
\end{itemize}

\hfill $\Box$

\begin{dfn}\label{d4.3.6}
\begin{itemize}
\item[(i)] A sign matrix satisfying (\ref{4.3.3}) is called an orthogonal column matrix (OCM).
\item[(ii)] A sign matrix satisfying (\ref{4.3.4}) is called
an almost orthogonal column matrix (AOCM).
\item[(iii)] A sign matrix $\mathbb{O}_{\a}\in {\cal M}_{\a\times \b}$ is called a largest OCM (LCOM), if it is an OCM with
maximum number of columns.
\item[(iv)] A sign matrix $\mathbb{U}_{\a}\in {\cal M}_{\a\times \b}$ is called a largest AOCM(LACOM), if it is an AOCM with
maximum number of columns.
\end{itemize}
\end{dfn}
%
%
%
%
%

\subsection{Constructing $\mathbb{O}$ and $\mathbb{U}$}

Assume the vertically expanded matrix $H_*\in {\cal M}_{m\times \a}$ is obtained as in (\ref{4.3.1}). Then
we consider how to construct  $\mathbb{O}$ (or $\mathbb{U}$), where
$$
\mathbb{O}\in {\cal M}_{t\times s},\quad (\mbox{or}~\mathbb{O}\in {\cal M}_{t\times s}),
$$
where
$$
t=\a-1.
$$
And we wish $s/t$ to be as large as possible.

\begin{lem}\label{l4.4.1} Assume $A$ is an OCM (or AOCM).
\begin{itemize}
\item[(i)] Replacing $\Col_j(A)$ by $-\Col_j(A)$, the resulting matrix is still an OCM (or AOCM).
\item[(ii)] Replacing $\Row_i(A)$ by $-\Row_i(A)$, the resulting matrix is still an OCM(or AOCM).
\item[(iii)] Doing a row (or column) permutation, the resulting matrix is still  an OCM(or AOCM).
\end{itemize}
\end{lem}

\noindent{\it Proof.} A simple computation shows that all the aforementioned operations do not change the coherence $\mu(A)$.
The conclusion follows.

\hfill $\Box$

In this subsection, denote by $A\sim B$,  if $B$ is obtained from $A$ by the transformations defined in Lemma \ref{l4.4.1}.

Assume $\Xi=\{\xi^1,\xi^2,\cdots,\xi^s\}$ is a set of $t$-dimensional  orthogonal vectors. Using Lemma \ref{l4.4.1}, we can, w.l.g., assume
$$
\xi^1:=\J_{t}\in \Xi.
$$
Assume $t$ is even, say $t=2t_2$, where $t_2\in \Z_+$ .  then, w.l.g., we can assume
$$
\xi^2=
\begin{bmatrix}
1\\-1
\end{bmatrix}
\otimes \J_{t_2}
\in \Xi.
$$
Now assume
$$
\xi^3=\begin{bmatrix}
\eta_1\\
\eta_2
\end{bmatrix}\in \Xi.
$$
Since $\xi^3$ is orthogonal with $\xi^1$, we have
$$
N_+(\eta_1)=N_-(\eta_2),\quad N_-(\eta_1)=N_+(\eta_2).
$$
Since $\xi^3$ is orthogonal with $\xi^2$, we have
$$
N_+(\eta_1)=N_+(\eta_2), \quad  N_-(\eta_1)=N_-(\eta_2).
$$
We conclude that
\begin{align}\label{4.4.1}
N_+(\eta_1)=N_-(\eta_1),\quad N_+(\eta_2)=N_-(\eta_2).
\end{align}
(\ref{4.4.1}) implies the following two facts:
\begin{itemize}
\item[(i)] If there exists $\xi^3$, then $t_2$ is an even number, and denoted by $t_2=2t_3$.
\item[(ii)]
$$
\eta_1,\eta_2\in \left\{\begin{bmatrix}1\\1\end{bmatrix}\otimes \J_{t_3}, \begin{bmatrix}1\\-1\end{bmatrix}\otimes \J_{t_3}\right\}.
$$
\end{itemize}
Hence, we can exactly have two more elements for $\Xi$, which are
$$
\xi^3=\begin{bmatrix}
\begin{pmatrix}
1\\-1
\end{pmatrix}\otimes \J_{t_3}\\
\begin{pmatrix}
1\\-1
\end{pmatrix}\otimes \J_{t_3}\\
\end{bmatrix};\quad
\xi^4=\begin{bmatrix}
\begin{pmatrix}
1\\-1
\end{pmatrix}\otimes \J_{t_3}\\
\begin{pmatrix}
-1\\1
\end{pmatrix}\otimes \J_{t_3}\\
\end{bmatrix}
$$
Continuing this procedure, the following result about $\mathbb{O}$ can be obtained.

As for $\mathbb{U}_{t}$, since we are not able to find a sign vector $x$ such that the coherence of $x$ and $\J_{t}$ is $\pm 1$.
  $\mathbb{U}_{t}=\mathbb{O}_{t}$. We, therefore, have the following result.

\begin{thm}\label{t4.4.2} Assume $t=2^pq$, where $q$ is an odd number.
\begin{itemize}
\item[(i)]  $\mathbb{O}_{t}$ can be obtained as
\begin{align}\label{4.4.2}
\mathbb{O}_{t}=\underbrace{\mathbb{O}_{2}\otimes \mathbb{O}_{2}\otimes \cdots\otimes \mathbb{O}_{2}}_p\otimes \J_q\in {\cal M}_{\a\times 2^p}.
\end{align}
where
$$
\mathbb{O}_2=\begin{bmatrix}1&1\\1&-1\end{bmatrix}.
$$
Moreover, $\mathbb{O}_{t}$ is unique up to column (or row) sign changes and column (or row) permutations.
\item[(ii)]
\begin{align}\label{4.4.3}
\mathbb{U}_{t}=\mathbb{O}_{t}.
\end{align}
\end{itemize}
\end{thm}

We conclude that

\begin{cor}\label{c4.4.3} Consider even $t$. When $t=2^p$, the  largest ratio $s/t$ can be obtained as
\begin{align}\label{4.4.4}
\max_{t}(s/t)=1.
\end{align}
\end{cor}

Next, we consider odd case, i.e., set $t=2k+1$.
Then $\mathbb{O}_{t}=\emptyset$. We construct $\mathbb{U}_{t}$.

Motivated by the case of even $t$, we may choose $t=2^p-1$.

\begin{prp}\label{p4.4.4} Consider odd $t$. When $t=2^p-1$, the ratio $s/t$ can be obtained as
\begin{align}\label{4.4.5}
\max_{t}(s/t)=\frac{2^p}{2^p-1},
\end{align}
which is slightly larger than $1$.
\end{prp}

\noindent{\it Proof.} Consider $\mathbb{O}_{2^p}$. Deleting the first row, the remainder becomes a $\mathbb{U}_{2^p-1}$.

\hfill $\Box$.

\begin{prp}\label{p4.4.5}

\end{prp}
\begin{prp}\label{pcs.3.3.5}
 The  $U\in \mathbb{U}_{2^p-1}$ generated from $\mathbb{O}_{2^p}$  is a maximum almost orthogonal matrix.
\end{prp}

 \noindent{\it Proof.}
 Assume
$$
U\in \mathbb{U}_{2^p-1}
$$
is generated from $\mathbb{O}_{2^p}$  by delete its first row. Then
$$
\mathbb{O}_{2^p}=
\begin{bmatrix}
\J^T_{2^p}\\
U
\end{bmatrix}.
$$
Hence,
$$
\langle \Col_i(U),\Col_j(U)\rangle=-1,\quad i\neq j.
$$
Note that
$$
\Col_1(U)=\J_{2^p-1},
$$
then
\begin{align}\label{cs.3.3.501}
\begin{array}{l}
N_+(\Col_j(U))=2^{p-1}-1,\\
N_-(\Col_j(U))=2^{p-1};\quad j\in [2,2^p].\\
\end{array}
\end{align}

We use contradiction to prove $U$ is the maximum almost orthogonal matrix. To this end, assume  $x\in \R^{2^p-1}$  is a sign vector, satisfying
 $\langle x, u\rangle=\pm 1$, $u\in \Col(U)$. i.e. $[U,x]\in \mathbb{U}_{2^p-1}$.

According to Lemma  \ref{lcs.3.3.1}, we can assume
$$
\langle x,\J_{2^p-1}\rangle=-1.
$$
Then $x$ satisfies (\ref{cs.3.3.501}). First,
$$
\langle x,u \rangle =-1,\quad \forall u\in \Col(U)
$$
is impossible. Otherwise,
$$
\begin{bmatrix}
\J_{2^p}&1\\
U&x
\end{bmatrix}\in \mathbb{O}_{2^p}.
$$
This contradicts the structure of maximum $\mathbb{O}_{2^p}$.
Hence, there is at least one  $y=\Col_j(U)$, $2\leq j\leq 2^p$, such that
\begin{align}\label{cs.3.3.502}
\langle x,y \rangle=1.
\end{align}
Since $y$ satisfies (\ref{cs.3.3.501}), we have
$$
\begin{array}{l}
|\{i\;|\; y_i=1,x_i=1\}|:=s,\\
|\{i\;|\; y_i=1,x_i=-1\}|=2^{p-1}-1-s,\\
|\{i\;|\; y_i=-1,x_i=1\}|:=t,\\
|\{i\;|\; y_i=-1,x_i=-1\}|=2^{p-1}-t.\\
\end{array}
$$
 $x$ also satisfies (\ref{cs.3.3.501}), hence
\begin{align}\label{cs.3.3.503}
s+t=2^{p-1}-1.
\end{align}
According to (\ref{cs.3.3.502}), we have
$$
s+2^{p-1}-t=2^{p-1}.
$$
That is
\begin{align}\label{cs.3.3.504}
s=t.
\end{align}
Using (\ref{cs.3.3.503}) and (\ref{cs.3.3.504}) yields
$$
s=t=\frac{2^{p-1}-1}{2},
$$
this is a contradiction because $s$  and $t$ are integers.

\hfill $\Box$

\begin{exa}\label{e4.4.6}
\begin{itemize}
\item[(i)] Consider $t=3$. Note that
$$
\mathbb{O}_4=\begin{bmatrix}
1&1&1&1\\
1&-1&1&-1\\
1&1&-1&-1\\
1&-1&-1&1\\
\end{bmatrix}.
$$
Then we have
$$
\mathbb{U}_3=\begin{bmatrix}
1&-1&1&-1\\
1&1&-1&-1\\
1&-1&-1&1\\
\end{bmatrix}\sim \begin{bmatrix}
1&1&1&1\\
1&-1&-1&1\\
1&1&-1&-1\\
\end{bmatrix}.
$$
\item[(ii)] Consider $t=7$. Using $\mathbb{O}_8$, we have
$$
\begin{array}{l}
\mathbb{U}_7=
\begin{bmatrix}
1&-1&   1&-1&  1&-1&  1&-1\\
1&  1& -1&-1&  1&  1&-1&-1\\
1& -1&-1&  1&  1&-1& -1&1\\
1&  1&  1&  1&-1&-1& -1&-1\\
1&-1&  1& -1&-1& 1& -1&1\\
1&  1&-1& -1&-1&-1&  1&1\\
1&-1&-1&   1&-1& 1&  1&-1\\
\end{bmatrix}\\
\sim \begin{bmatrix}
1& 1&  1& 1&  1& 1&  1& 1\\
1&-1&-1& 1&  1&-1&-1& 1\\
1& 1&-1&-1&  1& 1&-1&-1\\
1&-1& 1&-1&-1& 1&-1& 1\\
1& 1& 1&  1&-1&-1&-1&-1\\
1&-1&-1& 1&-1& 1&  1&-1\\
1& 1&-1&-1&-1&-1& 1& 1\\
\end{bmatrix}.
\end{array}
$$
\end{itemize}
\end{exa}

\begin{rem}\label{r4.4.6}
\begin{itemize}
\item[(i)]
If $t=2^p+1$, we may one arbitrary row to $\mathbb{O}_{2^p}$ to get a $\mathbb{U}_{2^p+1}\in {\cal M}_{2^p+1\times 2^p}$. But this one might not be the one with maximum ratio $s/t$. For instance, we have
$$
\mathbb{U}_5=
\begin{bmatrix}
1&1&1&1&1\\
1&1&1&-1&-1\\
1&1&-1&1&-1\\
1&-1&-1&-1&1\\
1&-1&1&1&-1\\
\end{bmatrix}\in {\cal M}_{5\times 5}.
$$
\item[(ii)] Unlike even case, so far we don't know how to construct $\mathbb{O}_{t}$ for general odd $t$.

\item[(iii)] Our conjecture is (\ref{4.4.5}) is the best ratio for odd $t$.

\item[(iv)] Using (\ref{4.3.101}) and (\ref{4.4.4}), we have that when $t=\a-1$ is even the best compression rate is $2$. If the above conjecture is correct, when $t$ is odd, the best compression rate is slightly higher than $2$.

\end{itemize}
\end{rem}

\section{Conclusion}

The purpose of this paper is to reveal some mathematical foundations for the STP-CS. First, the signal space is presented. The  STP based equivalence leads to a quotient space, which is the signal space. Second, a coordinate-free STP-CS is presented. It is also revealed that STP-CS has an advantage in $\ell_0$. Third, a systematic construction of BIBD-based sensing matrix is obtained.

\end{document}